\documentclass[12pt]{iopart}

\usepackage{graphicx}
\usepackage{iopams}  
\usepackage[usenames]{color}
\usepackage{bbm} % for blackboard bold math

\usepackage{aas_macros}

\newcommand*{\defeq}{\mathrel{\vcenter{\baselineskip0.5ex \lineskiplimit0pt
			\hbox{\scriptsize.}\hbox{\scriptsize.}}}=}
 \definecolor {darkgreen}{rgb}{0.2,0.7,0.2}

\begin{document}
\title[Analytic estimates for source confusion and parameter estimation errors]{Global analysis for the LISA gravitational wave observatory}

\author{Travis Robson and Neil Cornish}

\address{eXtreme Gravity Institute, Department of Physics, Montana State University, Bozeman, MT 59717}

\ead{travis.robson@montana.edu}
\ead{ncornish@montana.edu}

\begin{abstract}
The Laser Interferometer Space Antenna (LISA) will explore the source-rich milli-Hertz band of the gravitational wave spectrum. In contrast to ground based detectors, where typical signals are short-lived and discrete,  LISA signals are typically long-lived and over-lapping, thus requiring a global data analysis solution that is very different to the source-by-source analysis that has been developed for ground based gravitational wave astronomy. Across the LISA band, gravitational waves are both signals {\em and} noise. The dominant contribution to this so-called confusion noise (better termed unresolved signal noise) is expected to come from short period galactic white dwarf binaries, but all sources, including massive black hole binaries and extreme mass ratio captures will also contribute. Previous estimates for the galactic confusion noise have assumed perfect signal subtraction. Here we provide analytic estimates for the signal subtraction residuals and the impact they have on parameter estimation while for the first time incorporating the effects of noise modeling. The analytic estimates are found using a maximum likelihood approximation to the full global Bayesian analysis. We find that while the confusion noise is {\em lowered} in the global analysis, the waveform errors for individual sources are {\em increased} relative to estimates for isolated signals. We provide estimates for how parameter estimation errors are inflated from various parts of a global analysis.
\end{abstract}

\section{Introduction}
The recent discovery of gravitational waves~\cite{PhysRevLett.116.061102} and the outstanding success of the LISA pathfinder mission~\cite{PhysRevLett.116.231101} have given new life to the LISA mission. Building on decades of study, an updated LISA mission concept~\cite{LISA16} was recently submitted to address the European Space Agency's ``Gravitational Universe'' science theme with a launch scheduled for the early 2030's. The plan is to fly three identical spacecraft connected by six laser links forming a triangular detector with 2.5 million km long arms.

It has long been recognized that the LISA mission will suffer from ``an embarrassment of riches'', delivering data sets so packed with signals that extracting information about individual sources will require the development of unique data analysis techniques. Significant attention was given to this problem through the 2000's, culminating in a series of Mock LISA Data Challenges (MLDCs)~\cite{Arnaud:2006gm, Arnaud:2007vr, Arnaud:2007jy, Babak:2007zd, Babak:2008aa, Babak:2009cj} that produced some promising proof-of-principle solutions. The demise of the original LISA project in 2010 halted this effort, but work is now resuming following the re-birth of the mission. In addition to finding an implementable solution to the global analysis problem, there is also interest in producing reliable estimates for the science that can be achieved, including the number of sources of each type that can be resolved, and how well they can be characterized. A key input to these studies are estimates for the confusion noise from unresolved sources, as this adds to the instrument noise, and reduces the signal-to-noise ratio (SNR) of the resolved systems. These confusion noise estimates use variants of the idealized iterative source subtraction scheme introduced by Timpano {\it et al} \cite{PhysRevD.73.122001}. We recently applied this technique to produce confusion noise estimates~\cite{Cornish:2017vip} that were used in the design study for the new LISA mission concept~\cite{LISA16}. There are, however, several deficiencies with the simple confusion noise estimates: it assume that the confusion noise is stationary when in fact it oscillates with a 6 month period; it neglects the parameter estimation errors for the subtracted signals and the waveform residuals; and it also neglects the impact the removal has on other resolvable signals such as massive black holes. Here we generalize the Timpano {\it et al} \cite{PhysRevD.73.122001} method to account for the waveform residuals and the impact they have on the detection of other signals. Some of our results were derived previously by Cutler and Harms~\cite{PhysRevD.73.042001} in studies of foreground subtraction for the Big Bang Observer, but many results are new, including analytic estimates for power spectrum of the waveform residuals, incorporating the process of noise modeling, and the impact on parameter estimation for other sources. We find that the parameter estimation errors caused by other resolved signals are typically small compared to those due to instrument noise and unresolved signals. The exception to this rule is when two signals have very high overlap, such as sometimes occurs for galactic binaries with near identical orbital periods and sky locations~\cite{Crowder:2004ca}.

Electromagnetic observations have identified $\sim50$ galactic binaries with orbital periods that put their predicted gravitational signals in the LISA band~\cite{LISA16}. Those that rise above the noise are refer to as ``verification binaries''. Population synthesis models predict that there are far more detectable systems waiting to be discovered, though the estimates have been lowered in the past decade as on-going surveys have been used to re-calibrate the models~\cite{0004-637X-758-2-131,2007MNRAS.382..685R,2013MNRAS.429.2143C}. It is estimated that there are hundreds of millions of galactic binaries GBs emitting gravitational waves in our galaxy. In the mid-band of the LISA sensitivity, between $\sim 0.5-3$ mHz, gravitational waves from these systems are expected to dominate over instrument noise, with the unresolved component producing what is termed ``confusion noise''. 

There have been several previous attempts at estimating the galactic confusion noise~\cite{1990ApJ...360...75H,0004-637X-537-1-334,PhysRevD.73.122001,0004-637X-758-2-131,Cornish:2017vip}. To characterize the confusion noise one must first determine how many galactic binaries are resolvable, but in order to figure out which binaries are resolvable, one must already have an estimate for the noise. The ideal solution is to perform a global fit, {\it e.g.} a full Bayesian analysis that fits both resolvable sources and noise at the same time as done by Littenberg~\cite{PhysRevD.84.063009}. Unfortunately, this procedure is extremely computationally intensive, and more efficient techniques are needed if we want to consider a range of population models and detector configurations for design studies. To this end, Timpano {\it et al} \cite{PhysRevD.73.122001} developed an iterative subtraction scheme which starts with a simulated data set comprised of an instrument noise realization and the superposition of all gravitational waves produced by synthetic population of galactic binaries. The signal-to-noise ratio (SNR) of the GBs is calculated, and those above a specified threshold SNR are subtracted perfectly {\it i.e.} the true waveform is removed from the data stream using the simulated signal parameters. The noise estimate is updated after the bright signals are removed, and the SNRs of the remaining sources are re-computed. Those above the detection threshold are removed, and the whole process is repeated. It typically takes 5-6 iterations for the solution to converge. 

It is this assumption of perfect signal recovery we wish to address in this paper. In reality the instrument plus confusion noise realization will randomly perturb the estimated parameters for the resolvable systems, resulting in an inaccurate signal recovery. Here we use the Maximum-Likelihood approximation and Fisher information matrix to estimate the parameter errors and waveform residuals. 

The outline of this paper is as follows: In Section~\ref{sec:gal} we briefly review the galactic population model used to produce realizations of the LISA data used in the analysis.  Next, in Section~\ref{sec:ML} we provide a review of relevant Maximum-Likelihood methods and how they can be used to estimation the noise-induced errors in signal recovery and parameter estimation of the resolved GBs. In Section~\ref{sec:MLnoise} we extend the usual ML analysis to include noise spectral estimation, and in Section~\ref{sec:Bayes} we illustrate the relevance of the ML to a full Bayesian analysis using a simple model of a sinusoid in stationary Gaussian noise. In Section~\ref{sec:multiple} we apply the Maximum-Likelihood approach to the global fitting of multiple signals and drive expressions for how the interaction between the signals impacts waveform and parameter estimation errors. We conclude in Section~\ref{sec:confusion} by computing an improved estimate for the galactic confusion noise that takes into account parameter estimation errors in the bright source removal.

\section{Instrument and Galactic Population Models}\label{sec:gal}

Our galaxy simulations use realizations of the the Toonen {\it et al}~\cite{2012A&A...546A..70T} population model provided by Valeriya Korol and Gijs Nelemans . The space density of interacting white dwarf binaries is reduced by a factor of ten relative to earlier models in response to the findings of recent observational studies\cite{2007MNRAS.382..685R,2013MNRAS.429.2143C}. The population has  $\sim 26$ million systems with gravitational wave frequencies above 0.1 mHz. The signals from these systems are simulated using an improved version of the fast waveform generation algorithm of Cornish and Littenberg~\cite{PhysRevD.76.083006}.  The improved algorithm removes the need to sum over terms in the Fourier convolution by referencing the carrier frequency $f_0$ to the nearest integer multiple of the sample frequency, such that $f_0 = m/T_{\rm obs} +\delta f$, and absorbing the factor of $e^{2 \pi i \delta f t}$ into the slowly varying part of the signal. This removes the need for the sum in equation (A24) of Ref.~\cite{PhysRevD.76.083006}, and significantly speeds up the waveform generation. In our analysis we use the full set of first-generation time-delay interferometry (TDI)~\cite{ 0004-637X-527-2-814} variables $\tilde{X}(f), \tilde{Y}(f), \tilde{Z}(f)$ given in Ref.~\cite{PhysRevD.76.083006}, but when displaying results we show the more familiar Michelson-equivalent signals. To obtain the Michelson equivalent sensitivity we make use of the relation $S_{X} = 4 \sin^{2}\left(f/f_{*}\right)S_{M}$ where $S_{X}$ is the noise as seen in the TDI $X$ data channel and $S_{M}$ is the equivalent Michelson noise and $f_{*}=c/(2\pi L)$ is the transfer frequency. Our instrument noise model assumes white position noise $S_{p}(f)$ and colored acceleration noise $S_{a}(f)$ with spectral densities
\begin{eqnarray}
\hspace{-0.2in} S_{p}(f) &= 8.9 \times 10^{-23} {\rm m}^{2} {\rm Hz}^{-1} \nonumber \\
\hspace{-0.2in} S_{a}(f) &= 9\times 10^{-30}\left[1+\left(\frac{10^{-4}{\rm Hz}}{f}\right)^{2}+16\left(\frac{2\times10^{-5} {\rm Hz}}{f}\right)^{10}\right]{\rm m}^{2}{\rm s}^{-4}{\rm Hz}^{-1}.
\end{eqnarray}
Under the assumption that the noise levels are the same in each link, we can form the noise-orthogonal $\{A,E,T\}$  channels~\cite{Prince:2002hp}
\begin{eqnarray}
A &= \frac{1}{3}\left(2 X - Y - Z\right) \nonumber\\
E &= \frac{1}{\sqrt{3}}(Z-Y) \nonumber\\
T &= \frac{1}{3}\left(X+Y+Z\right) \, .
\end{eqnarray}
Below the transfer frequency $f_{*}$, where most signals are found, the $T$ channel is far less sensitive to gravitational waves, and does not contribute to our analysis.

\begin{figure}[htp]
	\begin{centering}
		\includegraphics[clip=true,angle=0,width=0.8\textwidth]{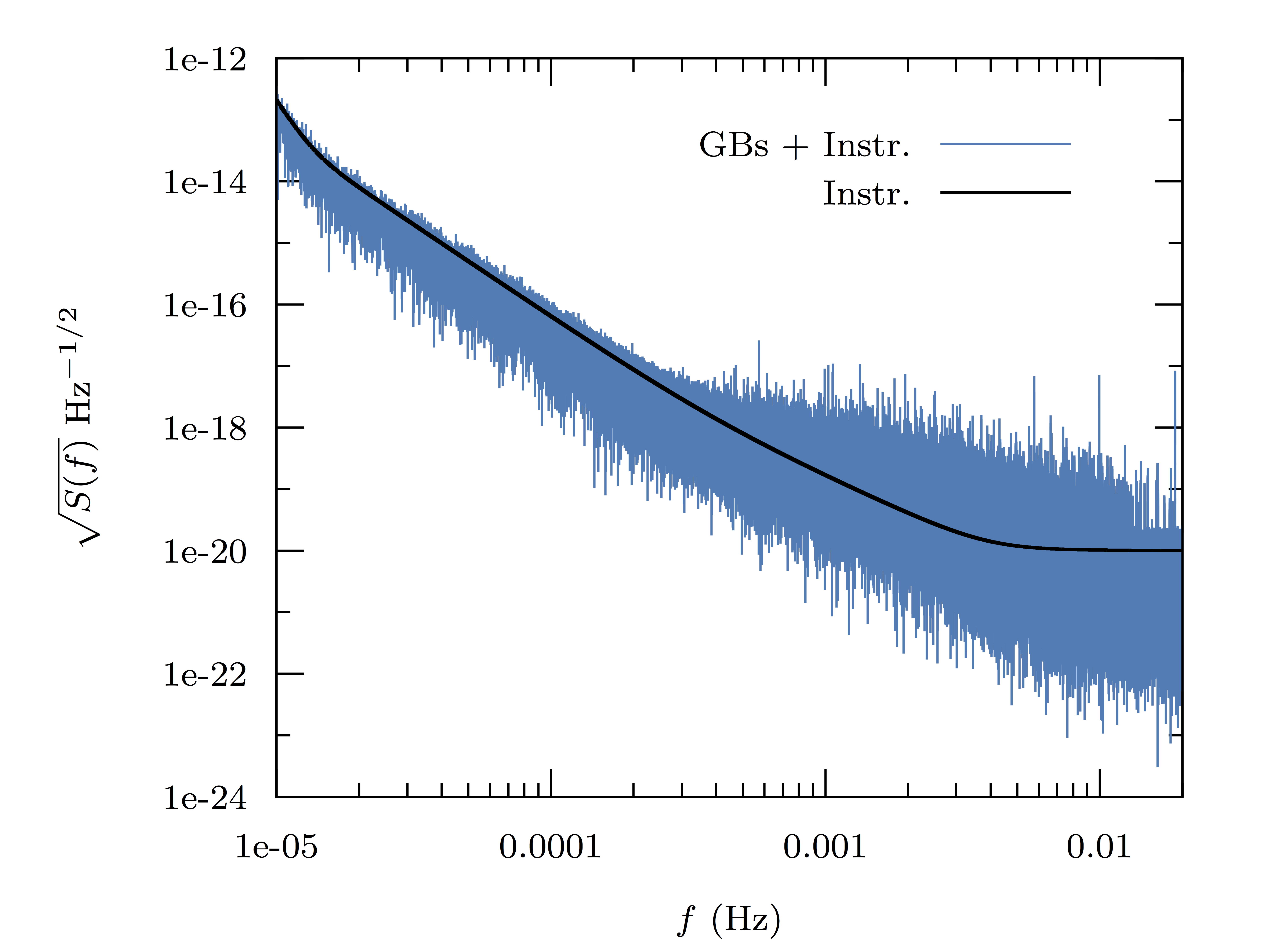} 
		\caption{\label{fig:galaxy_realization} A realization of the Michelson-equivalent strain spectral density in the $X$ channel, comprised of instrument noise and signals from our galactic binary population using a four year observation period. Note that signal from the galactic binaries dominates the instrument noise across the band $0.3 - 20$ mHz.}
	\end{centering}
\end{figure}

Figure \ref{fig:galaxy_realization} shows a realization of instrument noise (assumed to be Gaussian and stationary), combined with the signal from all of the relevant GBs in the population synthesis realization plotted as the Michelson-equivalent sensitivity. We see that the galactic foreground rises above the instrument noise across the frequency range $0.3-20$ mHz.

\section{Parameter estimation and waveform errors}\label{sec:ML}

In the high SNR regime, the likelihood is strongly peaked about the true model parameters, which allows for a Maximum-Likelihood analysis. Many useful results can be derived from a Taylor expansion of the likelihood about the true parameters. Here we provide a brief review of the Maximum-Likelihood (ML) approximation, deriving results for the parameter estimation errors and waveform residuals.  We follow with a discussion of how the ML analysis relates to a global Bayesian analysis.

\subsection{Maximum-Likelihood review}
Consider the simple case of data $\textbf{s}$ comprised of a gravitational wave signal $\textbf{h}_T = \textbf{h}(\vec{\lambda}_T)$ and stationary, Gaussian noise $\textbf{n}$. The likelihood of observing $\textbf{s}$ given the presence of a gravitational wave signal $\textbf{h}_T$ is then
\begin{equation}
p(\textbf{s}|\textbf{h}) = e^{- G/2} e^{-(\textbf{s}-\textbf{h}(\vec{\lambda})|\textbf{s}-\textbf{h}(\vec{\lambda}))/2} = e^{- G/2} e^{-\chi^2 /2} \,\, ,
\label{eq:likelihood}
\end{equation}
where
\begin{equation}
\left(\textbf{g}|\textbf{k}\right) \defeq 2\sum_{I=\{A,E\}}\int_{0}^{\infty} \frac{\tilde{g}_{I}(f)\tilde{k}_{I}^{*}(f) + \tilde{g}_{I}^{*}(f)\tilde{k}_{I}(f)}{S_{n,I}(f)}df \,\, ,
\label{eq:nwip}
\end{equation}
defines the noise-weighted inner product taken across all independent data channels, and the one-sided noise spectral density in channel $I$ is given by the expectation value
\begin{equation}
\mathbbm{E} [\tilde{n}_{I}(f) \tilde{n}_{J}^{*}(f^{\prime})] = \frac{1}{2}\delta(f-f^{\prime})S_{n,I}(f)\delta_{IJ} \,\, .
\label{eq:noise_psd}
\end{equation}
The noise $S_{n,I}(f)$ will include instrument noise and unresolved gravitational wave signals. The normalization factor $G$ is given by
\begin{equation}
G = \sum_{I=\{A,E\}}\int_{0}^{\infty} T \log\left[\pi T S_{n,I}(f)\right] df \, .
\label{eq:norm}
\end{equation}
The traditional derivation of the maximum likelihood solution assumes that the noise model $S_{n,I}(f)$ is known, and that the signal model, $\textbf{h}(\vec{\lambda})$ is close
to the true signal $\textbf{h}_T = \textbf{h}(\vec{\lambda}_T)$. The signal model is then Taylor expanded about the true parameters:
\begin{equation}
\textbf{h}(\vec{\lambda}) = \textbf{h}_T+\partial_i\textbf{h}_T\Delta \lambda^i + \mathcal{O}\left(\Delta\lambda^{2}\right),
\end{equation}
where $\Delta \vec{\lambda} =  \vec{\lambda} -  \vec{\lambda}_T$. The chi-squared in the likelihood can then be expanded as
\begin{eqnarray}
\fl \chi^2 = (\textbf{s}-\textbf{h}|\textbf{s}-\textbf{h}) \nonumber\\
    \fl   \quad = (\textbf{n}|\textbf{n}) - 2(\textbf{n}|\partial_i\textbf{h}_T)\Delta \lambda^i+(\partial_i\textbf{h}_T| \partial_j\textbf{h}_T) \Delta \lambda^i \Delta \lambda^j + {\cal O}(\Delta \lambda^3) \, .
\end{eqnarray}
The maximum likelihood solution is found by setting $\partial_i \chi^2 = 0$, which yields
\begin{equation}
\Delta \lambda^j = (\textbf{n}|\partial_i\textbf{h}_T)  \left(\Gamma^{-1}\right)^{ij} + \dots
\label{eq:dlambda}
\end{equation}
where 
\begin{equation}
\Gamma_{ij} = (\partial_i\textbf{h}_{T}| \partial_j\textbf{h}_{T})
\label{eq:fish}
\end{equation}
is the Fisher information matrix. Using the identity $\mathbbm{E} [ (\textbf{n}|\textbf{g})(\textbf{n}|\textbf{k})] = (\textbf{g}|\textbf{k})$ we find that the error covariance matrix is given to leading order in the signal-to-noise ratio by the inverse of the Fisher information matrix:
\begin{equation}
%C^{ij} = \rm I\!E [\Delta \lambda^{i} \Delta \lambda^{j}] = \left(\Gamma^{-1}\right)^{ij} + \mathcal{O}\left(SNR\right)^{-1} \,\, ,
C^{ij} = \mathbbm{E} [\Delta \lambda^{i} \Delta \lambda^{j}] = \left(\Gamma^{-1}\right)^{ij} + \mathcal{O}\left({\rm SNR}\right)^{-1} \,\, ,
\label{eq:param_covariance}
\end{equation}
where the SNR $\rho$ is given by $\rho^{2}(\textbf{h}) = (\textbf{h}|\textbf{h})$. See Vallisneri~\cite{PhysRevD.77.042001} for a more in depth presentation that discusses some of the potential pitfalls in using the Fisher Information matrix approximation  parameter error estimation. It is important to note that there are higher order corrections to the signal parameters and covariance matrix which appear in Cutler and Flanagan \cite{PhysRevD.49.2658} as equations (A31) and (A35). 

\subsection{Signal residuals}

We can use the maximum likelihood approximation to study noise induced errors in the parameter recovery and signal subtraction for galactic binaries. A closely related analysis was performed by Cutler and Harms~\cite{PhysRevD.73.042001} in the context of subtracting the signals from neutron stars to allow for the detection of a primordial stochastic background for the Big Bang Observer mission concept. We extend their analysis to include noise modeling, and derive new expressions for the impact the foreground removal has on parameter estimation for other sources such as massive black hole mergers and extreme mass ratio insprials (EMRIs).

\begin{figure}[htp]
	\begin{centering}
		\includegraphics[clip=true,angle=0,width=0.8\textwidth]{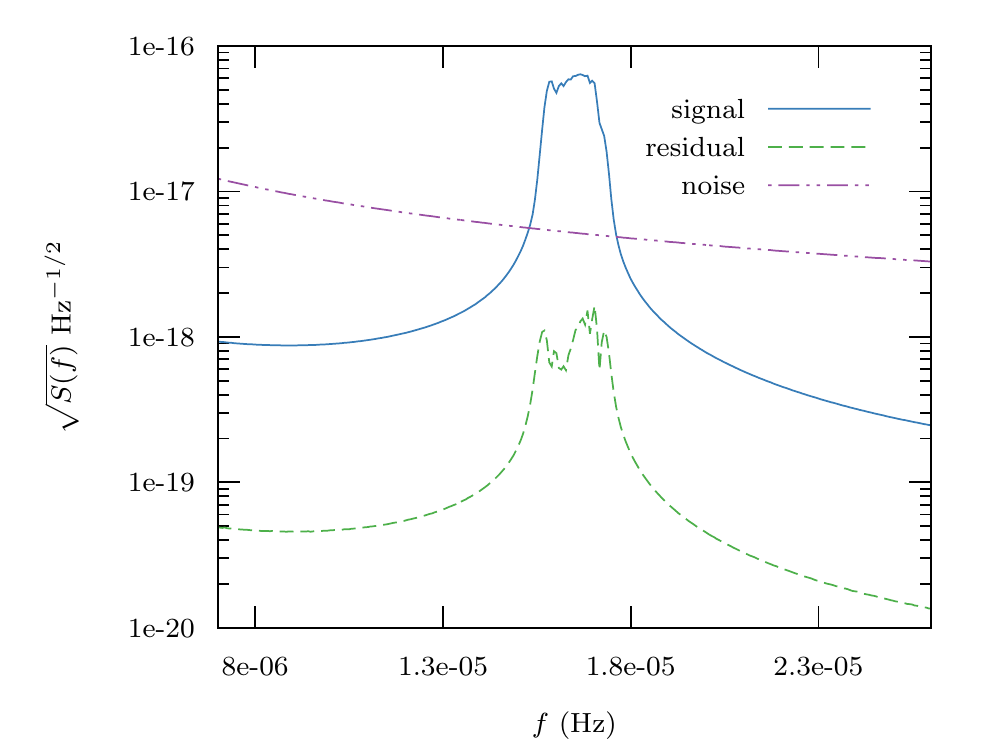} 
		\caption{\label{fig:sig_resid_noise}  This figure presents an example of the Michelson-equivalent strain spectral density for a signal (solid, blue) the corresponding waveform error (dashed, green) and the noise level (dot-dash, purple). The SNR of the signal was $\sim 110$ and the residual SNR was $\sim 5$. This signal results from one of the sources identified as bright after a 6 month observation run. Note the broadening that results from Doppler shift and frequency evolution.}
	\end{centering}
\end{figure}

The noise-induced parameter estimation errors (\ref{eq:dlambda}) result in waveform errors
\begin{equation}
\Delta \textbf{h} = \textbf{h}_T - \textbf{h} \simeq  - \partial_i \textbf{h}_T \Delta \lambda^{i} +\dots
\label{eq:dwave}
\end{equation}
An example of the observed signal from a galactic binary and the noise-induced subtraction residual is shown in Figure \ref{fig:sig_resid_noise}.
Note that the residual is below the reference noise level as the waveform error is down-weighted by the level of overlap between noise and parameter derivatives of the signal.

The waveform error has zero mean, $\mathbbm{E} [ \Delta \textbf{h} ] = 0$, and variance
\begin{equation}
\mathbbm{E} [\rho_{\Delta h}^{2}] \defeq \mathbbm{E} [(\Delta\textbf{h}|\Delta \textbf{h})] =   (\partial_i\textbf{h}_{T}| \partial_j\textbf{h}_{T}) \mathbbm{E} [\Delta \lambda^{i} \Delta \lambda^{j}] \approx  \Gamma_{ij}  \left(\Gamma^{-1}\right)^{ij}  = D \, .
\label{eq:dhsq}
\end{equation}
In the final step we have assumed that the error covariance matrix is approximated by the inverse of the Fisher matrix. The SNR of the residual depends only on the parameter dimension in the signal model and not upon the strength of the signal. Each term in the sum for $\Delta h$ is random walk induced by the noise realization, i.e. $\Delta h \sim n \sqrt{D}$ as there are $D$ terms. This means that $|\Delta \tilde{h}|^{2} \sim (n^{*}n)D$ and the inner product is weighted by the RMS noise resulting in a dependence only on model dimension. It can be shown that the variance of $\rho_{\Delta h}^{2}$ is $2D$ and the skew is $1/D^{2}$. The expectation value of the chi-squared can be written as
\begin{eqnarray}
\mathbbm{E}[\chi^2]&=  \mathbbm{E}[(\textbf{n}+\Delta\textbf{h}|\textbf{n}+\Delta\textbf{h})] \nonumber \\
&= N + \mathbbm{E} [\rho_{\Delta h}^{2}] + 2\mathbbm{E}\left[(\Delta\textbf{h}|\textbf{n})\right] = N - D\, ,
\label{eq:crev2}
\end{eqnarray}
where the last step follows from 
\begin{equation}
\mathbbm{E} [(\Delta \textbf{h}|\textbf{n})] = -\mathbbm{E} [(\textbf{h}_{,i}|\textbf{n})(\Gamma^{-1})^{ij}(\textbf{n}|\textbf{h}_{,j})]  = -D \,\, ,
\end{equation}
and $N$ is the number of data samples. We see that the signal residuals are anti-correlated with the noise, which results in a reduction in the chi-squared. Part of the noise gets absorbed by the signal model, which will ultimately result in a lowering of the confusion noise estimate relative to that found assuming perfect signal subtraction. Note that power spectrum of the residual $\textbf{s}-\textbf{h}$ has expectation value
\begin{eqnarray}
\fl \;  \mathbbm{E}[  (\textbf{s}(f)-\textbf{h}(f))(\textbf{s}(f')-\textbf{h}(f'))^* ] &= \frac{1}{2}S_{n,0}(f)\delta(f-f')  -  \partial_i\tilde{\textbf{h}}_T(f)\partial_j\tilde{\textbf{h}}^{*}_T(f') \left(\Gamma_0^{-1}\right)^{ij}\delta(f-f')   \nonumber \\
& =  \frac{1}{2} \left(S_{n,0}(f) - S_{\Delta \textbf{h}}(f)  \right) \delta(f-f')  \, ,
\end{eqnarray}
where
\begin{eqnarray}
\frac{1}{2}\delta(f-f^{'})S_{\Delta \textbf{h}}(f) & =  \partial_i\tilde{\textbf{h}}_T(f)\partial_j\tilde{\textbf{h}}^{*}_T(f') \left(\Gamma_0^{-1}\right)^{ij}\delta(f-f') \\
&= \mathbbm{E}[\Delta \tilde{\textbf{h}}(f)\Delta \tilde{\textbf{h}}^{*}(f')] 
\label{eq:sh}
\end{eqnarray}
is the power spectral density of $ \Delta \textbf{h}$. Note that we made use of $\mathbbm{E}\left[\tilde{\textbf{n}}(f) (\textbf{n}|\partial_{i}\textbf{h}_{T})\right] = \partial_{i}\tilde{\textbf{h}}_{T}(f)$.

\begin{figure}[htp]
	\begin{centering}
		\includegraphics[clip=true,angle=0,width=0.8\textwidth]{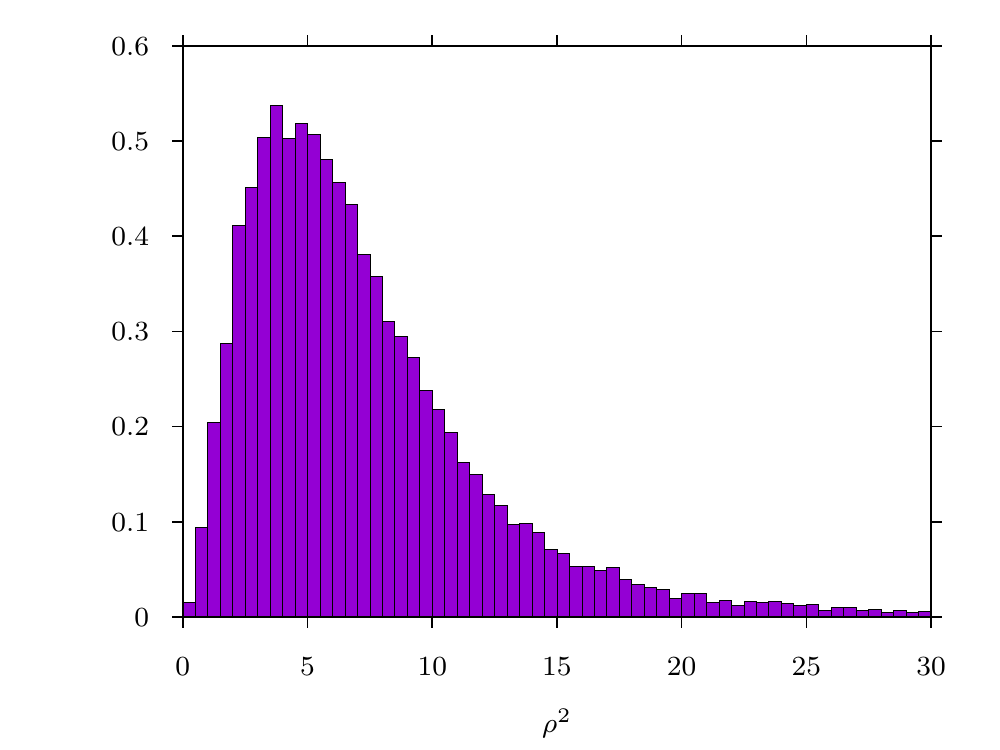} 
		\caption{\label{fig:residual_snr}  Histogram of the SNRs of the waveform errors resulting for the iterative subtraction scheme. The average value was 7.6.}
	\end{centering}
\end{figure}

Our waveform model for a galactic binary has $D=9$ parameters: the sky location $(\theta,\phi)$; inclination and polarization angles $(\iota,\psi)$; amplitude $A$; reference phase $\phi_0$ and reference gravitational wave frequency $f_0$; and first and second frequency derivatives $\dot f_0$ and  
$\ddot f_0$. For most galactic binaries the frequency derivatives are poorly constrained and the effective model dimension is closer to $D=7$. The relevant quantity for estimating which systems have detectable frequency evolution are the number of frequency bins of evolution, $\alpha = \dot f_0 T_{\rm obs}^2$ and $\beta = \ddot f_0 T_{\rm obs}^3$, and the SNR. Roughly speaking, frequency evolution through $\sim (7/{\rm SNR})$ bins is detectable~\cite{0004-637X-575-2-1030}, and to leading post-Newtonian order we have
 \begin{eqnarray}
  \alpha & = 1.5 \left( \frac{f_0}{4 \,  {\rm mHz}}\right)^{11/3}  \left( \frac{{\cal M} }{ 0.25 M_\odot }\right)^{5/3}  \left( \frac{ T_{\rm obs}}{4 \,  {\rm yrs}}\right)^{2}  \nonumber \\
   \beta & = 1.8 \left( \frac{f_0}{25 \,  {\rm mHz}}\right)^{19/3}  \left( \frac{{\cal M} }{ 0.25 M_\odot }\right)^{10/3}  \left( \frac{ T_{\rm obs}}{4 \,  {\rm yrs}}\right)^{3} \, .
   \label{eq:bins_evo}
  \end{eqnarray}
The chirp mass ${\cal M}$ has been scaled to the mode of the population distribution~\cite{Adams:2012qw}. From these expressions we see that only the loudest, most massive and highest-frequency systems will have a measurable second frequency derivative, and that most systems below 3 mHz will show no measurable frequency evolution at all. Including poorly constrained parameters in the model can lead to ill-conditioned Fisher matrices with badly behaved inverses. One solution is to reduce the model dimension by eliminating parameter $\lambda_i$ whenever the inner
product $(\partial_i\textbf{h}_{T}| \partial_i\textbf{h}_{T})$ drops below some threshold. An alternative solution is to replace the Fisher matrix $\boldsymbol  \Gamma$ with a matrix formed from the augmented Fisher matrix, ${\bf K}$, which includes derivatives of the priors (see Section~\ref{sec:Bayes} for details). We adopt the latter approach and include Gaussian priors on $\alpha,\beta$ centered on zero with width $\sigma = 10$. The priors condition the matrix and when the parameters are poorly constrained by the data, have the effect of reducing the model dimension.
Using the more stable approximation $\mathbbm{E} [\Delta \lambda^{i} \Delta \lambda^{j}] \approx \left(K^{-1}\right)^{ij}$, yields $\mathbbm{E} [\rho_{\Delta h}^{2}] \approx D_{\rm eff}$, where $D_{\rm eff}$ is the effective dimension of the model, defined by the number of parameters that have posterior distributions that differ measurably from their priors (a notion that can be made precise using the Kullback-Leibler divergence).

Figure \ref{fig:residual_snr} shows a histogram of square SNRs of the waveform residuals for the galactic binaries that are deemed detectable by the iterative subtraction scheme discussed in Section~\ref{sec:confusion}. The average value for there residual ${\rm SNR}^2$ of 7.6 is less than the full model dimension $D=9$, and consistent with our estimate for the effective dimension.

\section{Maximum-Likelihood approximation with noise estimation}\label{sec:MLnoise}

The standard treatment of the maximum likelihood expansion assumes that the noise spectrum is known. If the detectable gravitational wave signals are infrequent and short-lived, as is currently the case for compact binary mergers in LIGO, then ``off-source'' data from times where no detectable signals are present can be used to estimate the noise spectrum. These estimates will include instrument noise and unresolved gravitational wave signals. The option of making off-source estimates will not be available for LISA, and the noise spectrum will have to be inferred along with the signal model. Our derivation we will assume that we have a parameterized model for the noise, such as the cubic spline model used by the BayesLine algorithm~\cite{Littenberg:2014oda}.

To get an understanding for how noise modeling impacts the maximum likelihood calculation, consider a simple example with zero mean, additive, white Gaussian noise and $N$ data samples with likelihood
\begin{equation}
p(\textbf{s}|\textbf{h}) = \prod_{k=1}^N \frac{1}{\sqrt{2 \pi \sigma^2}}\,  e^{-(s_k - h_k)^2/(2 \sigma^2)} \, .
\label{eq:likelihood2}
\end{equation}
The un-perturbed ($\textbf{h} = 0$) noise level is given by sample variance
\begin{equation}
\sigma_0^2 = \frac{1}{N} \sum_{k=1}^{N} n_k^2 \, .
\label{eq:sigma0}
\end{equation}
We could expand $\sigma^2$ about the theoretical variance, $\sigma_*^2$, but it is simpler to expand $\sigma^2$ about the sample variance: $\sigma^2 = \sigma_0^2 + \Delta \kappa$ so that
$\Delta \kappa = 0$ when $\textbf{h} = 0$. The typical difference between the sample variance and the theoretical variance will be by an amount that scales as the standard deviation of the sample variance,
$\Delta \sigma_0^2 = \sqrt{2} \sigma_*^2/\sqrt{N}$. The log likelihood can be expanded:
\begin{eqnarray}
\log p(\textbf{s}|\textbf{h})  & = & \mbox{const} +  \left(1 - \frac{\Delta \kappa}{\sigma_0^2}\right)\left((\textbf{n}|\partial_i\textbf{h}_T)_0 \Delta \lambda^i-\frac{1}{2}(\partial_i\textbf{h}_T| \partial_j\textbf{h}_T)_0 \Delta \lambda^i \Delta \lambda^j  \right)  \nonumber \\
 && -\frac{N}{2} \left( 1+  \frac{\Delta \kappa^2}{2 \sigma_0^4} \right) + \dots \, ,
\label{eq:likelihoodX}
\end{eqnarray}
where the notation $(a\vert b)_0$ denotes that the noise weighted inner product is taken with respect to the un-perturbed noise level $\sigma_0^2$. Setting
$\partial_{\Delta \kappa} \log p(\textbf{s}|\textbf{h}) = 0$ and  $\partial_{\Delta \lambda^k} \log p(\textbf{s}|\textbf{h}) = 0$, yields the ML solution
\begin{eqnarray}
\Delta \lambda^j &=& (\textbf{n}|\partial_i\textbf{h}_T)_0  \left(\Gamma_0^{-1}\right)^{ij}  \nonumber \\
\Delta \kappa &=& - \frac{\sigma_0^2}{N} (\textbf{n}|\partial_i\textbf{h}_T)_0 (\textbf{n}|\partial_j\textbf{h}_T)_0 \left(\Gamma_0^{-1}\right)^{ij} \, .
\label{eq:ML2} 
\end{eqnarray}
We see that the leading order ML solution for the signal parameters is unchanged from the fixed noise case. The updated noise estimate $\sigma^2_{\rm ML} = \sigma^2_{0}+\Delta \kappa$ is {\em lowered} relative to its true value, as can be seen by taking the expectation value
\begin{equation}
\mathbbm{E} [ \sigma^2] =  \sigma^2_{0}\left( 1 - \frac{D}{N} \right) \, .
\label{eq:sigX}
\end{equation}
While the ML waveform removes some of the noise, this is now accounted for in the ML estimate for the noise, such that the expected value of the chi-squared is again just the dimension of the data: $\mathbbm{E} [  \chi^2 ] = N$. 

From expanding the likelihood around the signal and noise parameters $\vec{\eta} =\lbrace\vec{\lambda},  \kappa \rbrace$ and maximizing the likelihood obtains the form

\begin{equation}\label{eq:fullfish}
p(\textbf{s}|\Delta \vec{\eta}) = \frac{1}{\sqrt{2\pi \mbox{ det}\boldsymbol{\Gamma}^{-1}}}e^{-\frac{1}{2}\Gamma_{\mu \nu}\Delta \eta^{\mu} \Delta \eta^{\nu}}
\end{equation}
where $\Gamma_{\mu \nu} =-\partial_{\mu}\partial_{\nu}\log p(\textbf{s}|\textbf{h})|_{\scriptsize\mbox{max}}$. Note that we are using Greek indicies to denote the entire collection of parameters. One can read off the Fisher matrix from the log-likelihood:

\begin{equation}
\boldsymbol{\Gamma} = \left(
\begin{array}{cc}
	\Gamma_{0,ij} & (\textbf{n}|\partial_j\textbf{h}_T)_0/\sigma_{0}^{2} \\
	(\textbf{n}|\partial_i\textbf{h}_T)_0/\sigma_{0}^{2} & 2 \sigma_{0}^{4}/N 
\end{array}  \right) \, ,
\end{equation}
where $\Gamma_{0,ij}$ is the Fisher matrix obtained from the signal-only ML analysis discussed in the previous section. The full Fisher matrix can be inverted by recognizing that the off-diagonal terms $(\textbf{n}|\partial_j\textbf{h}_T)_0/\sigma_{0}^{2}$ are small  compared to the block diagonal terms.  We find that the variances for the signal parameters are inflated:

\begin{equation}
\left(\Gamma^{-1}\right)^{ii} \approx \left(\Gamma^{-1}_{0}\right)^{ii} + \frac{2}{N}\left(\Gamma^{-1}_{0}\right)^{il}\left(\Gamma^{-1}_{0}\right)^{ki}(\textbf{n}|\partial_l\textbf{h}_T)_0(\textbf{n}|\partial_k\textbf{h}_T)_0 \, .
\end{equation}
On average $\left(\Gamma^{-1}\right)^{ij} \rightarrow \left(\Gamma^{-1}_{0}\right)^{ij} \left(1+2/N\right)$ where in the limit of large $N$ we obtain the original Fisher matrix. The signal model parameter variances are inflated by covariances with the noise model parameters as they both attempt to capture pieces of the signal. Note that covariance of the parameter shifts  $\Delta \lambda^{j}$ and $\Delta \kappa$ from their true values, as computed in (\ref{eq:ML2}), does not equal the inverse of the Fisher matrix,  $\mathbbm{E}\left[\Delta \eta^{\nu}\Delta\eta^{\mu}\right] \neq (\Gamma^{-1})^{\mu\nu}$. This is because the noise modeling changes the shape of the likelihood, and not just the location of the peak. However, we see from (\ref{eq:fullfish}) that $\Gamma^{-1}$ does indeed describe the parameter covariances. 

The ML expansion for a colored noise model is considerably more involved, and we relegate the details to Appendix A. To keep the notation simple we suppress the sum over data channels.
Introducing the parameterized noise model
\begin{equation}
S_{n}(f;\vec{\theta}) = S_{n,0}(f) + \Delta \theta^{a}\partial_a S_{n}(f)  + \frac{1}{2} \Delta \theta^{a}\Delta \theta^{b}  \partial_a \partial_b S_{n}(f) +\dots \, ,
\label{eq:noisemod}
\end{equation}
(noise model derivatives are evaluated at the ML values after differentiation) where $S_{n,0}(f)$ is some smooth estimate of the instrument noise and unresolved signals (assuming perfect subtraction of the resolvable signals), we find the 
leading-order solution for the signal parameters has the same form as in (\ref{eq:ML2}), while the noise model parameters are given by
\begin{eqnarray}
\Delta \theta^{j} \approx &\left[ \int \frac{S_{n,i}S_{n,j}}{S_{n,0}^{2}}\left(\frac{T S_{n,0}- 4 \tilde{n}^{*}\tilde{n}}{S_{n,0}}\right)df - \int \frac{S_{n,ij}}{S_{n,0}}\left(\frac{T S_{n,0}- 2 \tilde{n}^{*}\tilde{n}}{S_{n,0}}\right)df \right]^{-1} \nonumber \\
& \times \Bigg[\int \frac{S_{n,i}}{S_{n,0}}\left(\frac{T S_{n,0}- 2 \tilde{n}^{*}\tilde{n}}{S_{n,0}}\right)df + 2(\textbf{n}|\partial_{a}\textbf{h}_{T})_{i}(\textbf{n}|\partial_{b}\textbf{h}_{T})_{0}\left(\Gamma_{0}^{-1}\right)^{ab} \Bigg. \nonumber \\
&\,\,\,\,\,\,\,\, \Bigg. - (\partial_{a}\textbf{h}_{T}|\partial_{b}\textbf{h}_{T})_{i}(\textbf{n}|\partial_{c}\textbf{h}_{T})_{0}(\textbf{n}|\partial_{d}\textbf{h}_{T})_{0} \left(\Gamma_{0}^{-1}\right)^{ac}\left(\Gamma_{0}^{-1}\right)^{bd}  \Bigg] \, .
\label{eq:noisepar}
\end{eqnarray}
The notation $(x|y)_a$ defines the inner product
\begin{equation}
(x|y)_a=  4 \int \frac{ ( \tilde{x} \tilde{y}^* + \tilde{x}^* \tilde{y}) }{ S_{n,0} }  \frac{\partial_a S_{n}}{ S_{n,0}} df \, .
\end{equation}
The integrals with the factor $(T S_{n,0}- 2 \tilde{n}^{*}\tilde{n})/S_{n,0}$ accounts for the difference between the theoretical noise model and fluctuation seen in a particular noise realization,  i.e. the difference between $\sigma_{*}^{2}$ and $\sigma_{0}^{2}$ in the white noise toy model, as evidenced by its expectation value vanishing. Neglecting this difference and considering the expectation value of $\Delta \theta^{j}$ we obtain the simplification 
\begin{eqnarray}
\mathbbm{E}\left[\Delta \theta^{j}\right] \approx &- \left[ \int T  \frac{S_{n,i}S_{n,j}}{S_{n,0}^{2}}df \right]^{-1} (\partial_{a}\textbf{h}_{T}|\partial_{b}\textbf{h}_{T})_{i}\left(\Gamma_{0}^{-1}\right)^{ab}  \,\, .
\end{eqnarray}
The white-noise case (\ref{eq:ML2}) can be recovered by setting $S_{n,0}= \frac{2}{T} \sigma_0^2$ and $\partial_a S_{n}(f) = \frac{2}{T}$, so that $ \left[ \int T  \frac{S_{n,i}S_{n,j}}{S_{n,0}^{2}}df \right]^{-1} = \sigma_0^4/N$, and $(x|y)_a= (x|y)_0/\sigma_0^2$. 

We can now  compute the expectation value of the noise perturbation $\Delta S_n(f) =  \Delta \theta^{a}\partial_a S_{n}(f) +\dots$
\begin{equation}
\mathbbm{E}[ \Delta S_n(f) ] =  -\left[ \int T  \frac{S_{n,a}S_{n,b}}{S_{n,0}^{2}}df \right]^{-1}   \left(\Gamma_0^{-1}\right)^{ij}  (\partial_i\textbf{h}_T |\partial_j\textbf{h}_T)_a \partial_b S_{n}(f) \, .
\end{equation}
Note that the perturbation to the noise model is negative, as it must be given that the signal model absorbs some of the noise.  One would expect that $\Delta S_{n}(f)$  should be a smoothed representation of $-S_{\Delta h}(f)$ for an effective noise model, mopping up errors introduced by the signal ML.

Similar to the white noise model above we may obtain the signal model variances for a general noise model by making the same appeals to neglecting differences between the theoretical and sample variance and averaging over many noise realizations

\begin{eqnarray}
	\left(\Gamma^{-1}\right)^{ii} \approx \left(\Gamma_{0}^{-1}\right)^{ii} + 2 \left(\Gamma_{0}^{-1}\right)^{im}\left(\Gamma_{0}^{-1}\right)^{ni}&(\partial_{m}\textbf{h}_{T}|\partial_{n}\textbf{h}_{T})_{ab}\nonumber \\
	&\times \left(\int_{0}^{\infty}T\frac{S_{n,a}S_{n,b}}{S_{n,0}^{2}}df \right)^{-1} \, .
\end{eqnarray}
An effective noise model would minimize the factors $(\partial_{m}\textbf{h}_{T}|\partial_{n}\textbf{h}_{T})_{ab}$ such that $\left(\Gamma^{-1}\right)^{ii} \rightarrow \left(\Gamma_{0}^{-1}\right)^{ii}$. We can turn this into a more useful expression by taking advantage of the compact (in the frequency domain) nature of the GB signal and assume the noise PSD is roughly constant

\begin{eqnarray}
\left(\Gamma^{-1}\right)^{ii} \approx \left(\Gamma_{0}^{-1}\right)^{ii}\left(1+\frac{2}{T\Delta f}\right) \, ,
\label{eq:cov_envo}
\end{eqnarray}
where $\Delta f$ is the bandwidth of the signal such that $T\Delta f$ is the number of frequency bins the GB spans. For sources that span many frequency bins such that the noise PSD cannot be assumed to be constant this serves as an upper limit for the increase in the parameter errors. Note that other terms exist when considering covariances of the signal model. Again, we see that when the source occupies a large bandwidth we recover the variances for when the noise is known. For a $3$ mHz source that experiences a $\Delta f \approx 0.6 \times 10^{-6}$ Hz Doppler shift spreading due to LISA's orbital motion, the parameter variances will be inflated by $10\%$ for a one year observation span, dropping to $3\%$ after four years.

\section{Relating Bayesian Inference and Frequentist Maximum Likelihood Estimation}\label{sec:Bayes}

The LISA data will include overlapping signals from an unknown number of sources of different types. Bayesian inference provides a powerful and flexible framework for inferring the number and properties of the resolvable sources. In addition to the ontological differences between the Bayesian and Frequentist approach to statistical inference - Bayesian inference considers the data to be known and the signal parameters to be uncertain while Frequentist inference considers the  signal parameters to be fixed and the data to be uncertain - Bayesian inference typically integrates over uncertainty (marginalization), while Frequentist analysis employs maximization. Despite these differences, the maximum likelihood analysis we have described can be used to estimate results from Bayesian inference by way of a Taylor expansion of the posterior distribution $p(\vec{\lambda}\vert \textbf{s}) = p( \textbf{s} \vert \vec{\lambda}) p(\vec{\lambda}) /p( \textbf{s})$. Expanding about the mode of the posterior distribution (also known as the  maximum a posteriori probability (MAP) estimate, $\vec{\lambda}_{\rm MAP}$), we have the quadratic approximation
\begin{equation}
p(\vec{\lambda}\vert \textbf{s}) \simeq (2\pi)^{-D/2} ({\rm det} \boldsymbol K)^{1/2} e^{-\frac{1}{2} K_{ij} \Delta \lambda^i \Delta \lambda^j} \, ,
\label{eq:map}
\end{equation}
where $\Delta  \vec{\lambda} =  \vec{\lambda} -  \vec{\lambda}_{\rm MAP}$ and
\begin{equation}
K_{ij}  = -\partial_i\partial_j   \log\left(p( \textbf{s} \vert \vec{\lambda}) p(\vec{\lambda})\right) \bigg\vert_{\rm MAP}  \, . 
\label{eq:Kappa}
\end{equation}
When the likelihood is more informative than the prior,  $\vec{\lambda}_{\rm MAP} \approx  \vec{\lambda}_{\rm ML}$ and ${\bf K}$ is well approximated by the Fisher information matrix $\boldsymbol  \Gamma$, though even small contribution from the derivatives of the log prior can have a an important stabilizing effect on ${\bf K}$.

\begin{figure}[htp]
	\begin{centering}
		\includegraphics[clip=true,angle=0,width=0.7\textwidth]{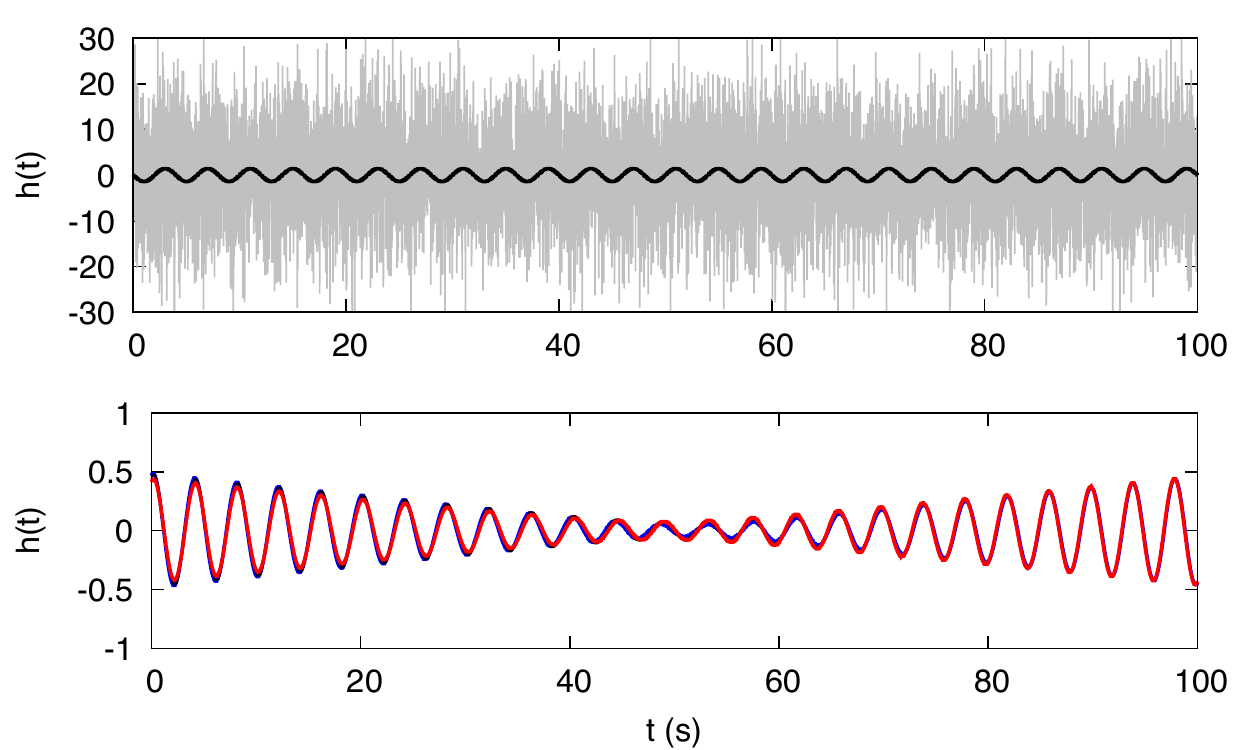} 
		\caption{\label{fig:sinusoid}  The upper panel shows simulated data (grey) and signal (black) for the simple sinusoid model. The lower panel shows the waveform residuals computed using the analytic maximum likelihood (red), and the numerical maximum a posteriori (blue) and mean (black) values from a MCMC analysis.}
	\end{centering}
\end{figure}

To illustrate the relationship between the maximum likelihood analysis and Bayesian inference we produced simulated data consisting of stationary, Gaussian white noise with variance $\sigma^2$ and a sinusoidal signal $h(A,f_0, t_0,\phi_0) = A \cos( 2 \pi f_0 (t-t_0)+\phi_0)$. 
We held the phase parameter $\phi_0 = \pi$ fixed in the analysis as otherwise there is a near perfect degeneracy between the time offset $t_0$ and the phase offset $\phi_0$, which significantly complicates the analysis. The simulated data consisted of $N= 10^4$ evenly spaced samples spanning $T= 100$ seconds, with $A=\sqrt{2}$, $f_0=0.25$ Hz, $t_0=1$ second and $\sigma^2 = 100$. The noise level was set to yield a signal-to-noise ratio of ${\rm SNR = 10}$.  A plot of the simulated data and signal are shown in the upper panel of Figure~\ref{fig:sinusoid}.

\begin{table}[ht]
\caption{Parameter Error Estimates}
\centering
\begin{tabular}{c c c c c c}
\hline\hline
Parameter & $\sigma_\Gamma$ & $\sigma_{\rm MCMC}$ & $\Delta \lambda_{\rm ML}$  & $\Delta \lambda_{\rm MAP}$ &  $\Delta \lambda_{\rm mean}$ \\ [0.5ex]
\hline
$A$ & 0.141 & 0.135 & -0.078 &  -0.048 & -0.067 \\
$f_0$ & 0.00055 & 0.00054 & 0.0010 & 0.0011 & 0.0010 \\
$t_0$ & 0.125 & 0.123 & 0.195 & 0.214 & 0.217 \\ [1ex]
\hline
\end{tabular}
\label{table:par}
\end{table}

\begin{figure}[htp]
	\begin{centering}
		\includegraphics[clip=true,angle=0,width=\textwidth]{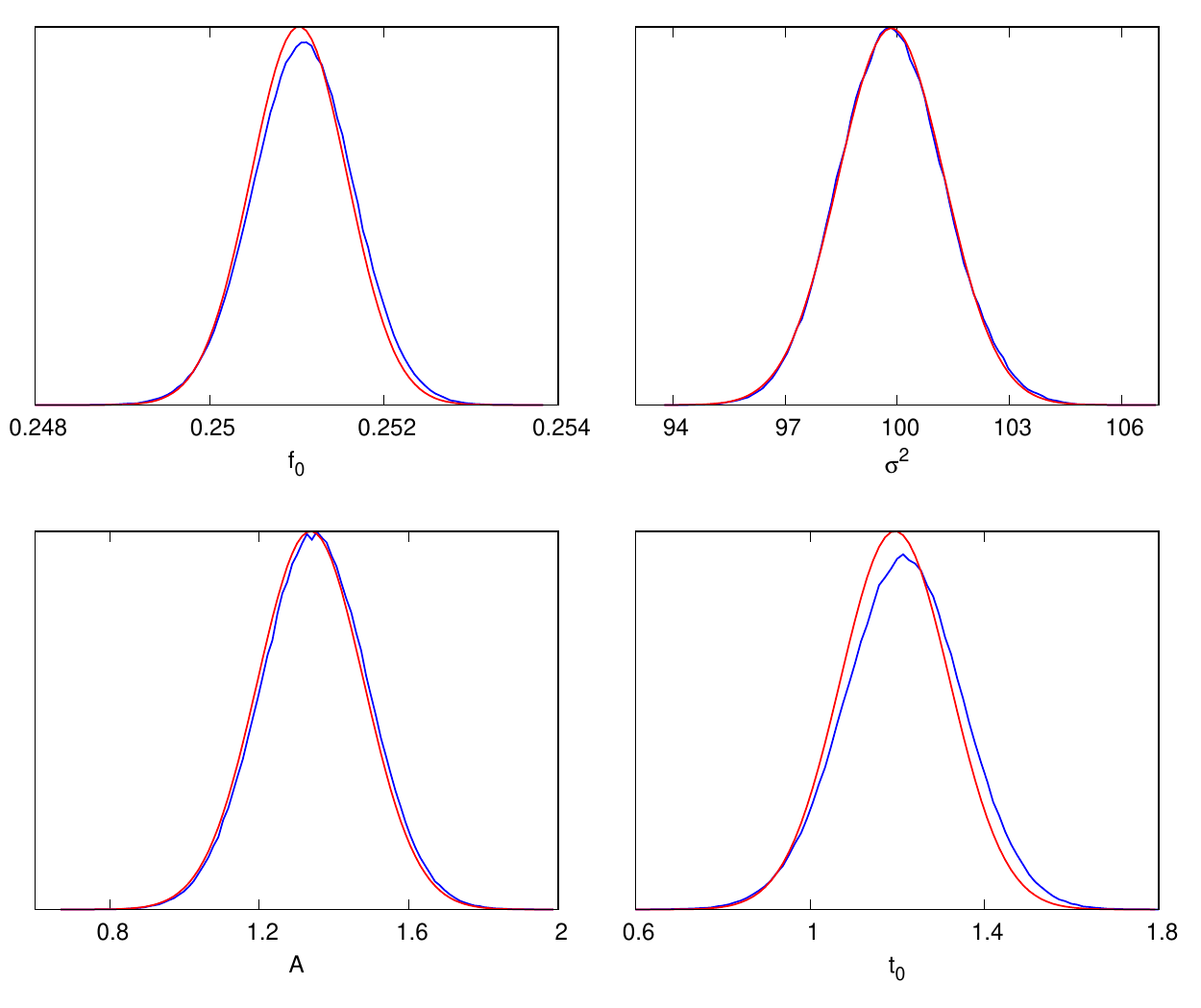} 
		\caption{\label{fig:post}  Marginalized posterior distributions for the signal and noise parameters from a MCMC simulation (blue) and the Gaussian ML approximation (red).}
	\end{centering}
\end{figure}

For the Bayesian analysis we assumed uniform priors on the signal and noise model parameters $(A,f_0, t_0, \sigma^2)$ across a range that was much wider than the predicted statistical errors so that posterior distribution and the likelihood were effectively identical. A Markov Chain Monte Carlo (MCMC) simulation was used to compute the mean and variance of the signal parameters and the waveform error, while (\ref{eq:ML2}) and (\ref{eq:fullfish}) were used to estimate the parameter shifts at maximum likelihood and the variances. The MCMC and ML derived values for the parameter shifts and standard deviations are listed in Table~\ref{table:par} for a particular noise realization. The marginalized posterior distributions for the parameters are compared to the predictions of the Gaussian approximation (\ref{eq:map}) in Figure \ref{fig:post}. The agreement between the ML and MCMC seen in this example was typical of what we found when repeating the analysis for different noise realizations. That is not to say that the Gaussian approximation will be this accurate in more realistic settings where the noise is more complicated and the parameters are highly correlated~\cite{PhysRevD.77.042001}, but it does provide useful order-of-magnitude estimates in most situations.

The displacement of the parameters from their true values is related to the waveform error $\Delta h$ shown in the lower panel of Figure~\ref{fig:sinusoid} which displays the ML, MAP and mean waveform errors. Note that the mean waveform error is found by averaging the waveform errors, and {\em not} by using the mean parameter values to compute a waveform. 

\begin{figure}[htp]
	\begin{centering}
		\includegraphics[clip=true,angle=0,width=0.7\textwidth]{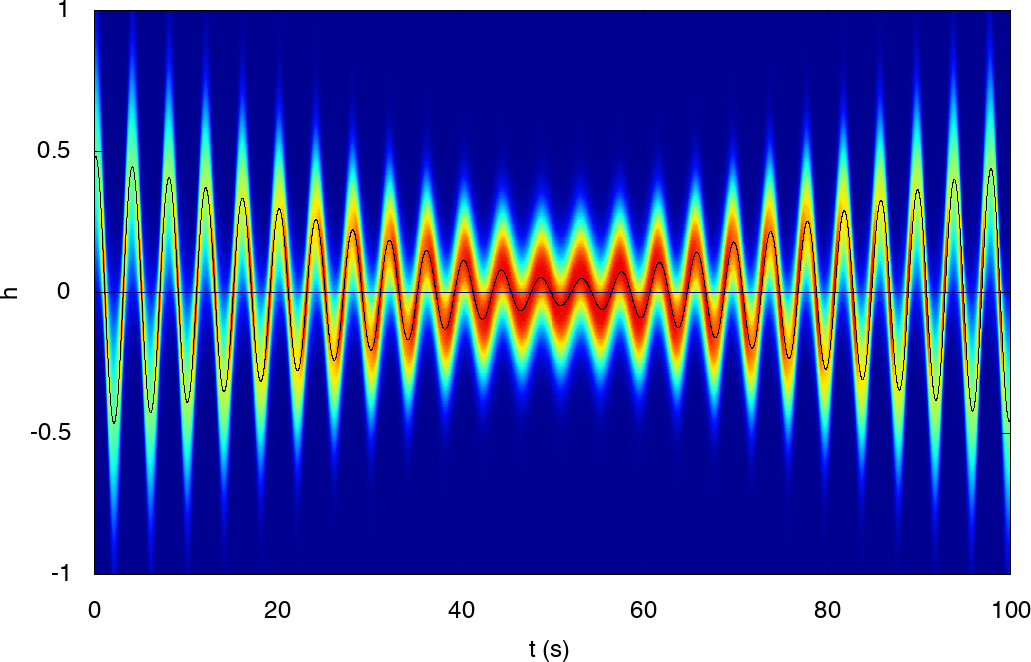} 
		\caption{\label{fig:dh}  The posterior distribution for the waveform error computed from the MCMC analysis using a heat map to indicate the posterior weight, compared to the MAP point estimate (thin black line).}
	\end{centering}
\end{figure}

The good agreement between the maximum likelihood and mean waveform residual hides a key difference between the frequentist and Bayesian analyses: in the Bayesian global fit the waveform uncertainties are marginalized over, while the frequentist analysis uses point estimates. Rather than subtracting a particular point estimate for each signal from the data, the Bayesian approach subtracts a range of estimates for each signal such that the residual is consistent with the noise model. This procedure is illustrated in Figure~\ref{fig:dh} for the sinusoid signal model, where the waveform residuals from each iteration of the MCMC analysis are used to produce a histogram of the residual at each time sample. Also shown in
Figure~\ref{fig:dh} is the MAP point estimate for the waveform residual. Notice that the full posterior distribution for the waveform residuals has significant spread about the point estimate.

\section{Multiple Sources}\label{sec:multiple}

The LISA data will contain many signals that partially overlap in both time and frequency. Extracting information about these signals necessitates finding a global solution for all signals that can be resolved - that is, signals that are both sufficiently loud and sufficiently distinct to be individually identified. The full LISA data stream can be written as 
\begin{equation}
\textbf{s} = \textbf{H} + \textbf{n} \, ,
\label{eq:LISAdata}
\end{equation}
where $\textbf{H} =  \sum \textbf{h}_{T}$ denotes the sum total of all gravitational wave signals, which can be further separated into a resolved, $\textbf{H}_R$, and un-resolved $\textbf{H}_U$, component. The unresolved component is often referred to as ``confusion noise'', the largest component of which is expected to come from white dwarf binaries 
in our galaxy. The ML analysis for single sources can be applied to multiple sources by replacing $\textbf{h}$ with $\textbf{H}_R$, and by replacing $\textbf{n}$ by  $\textbf{n}+\textbf{H}_U$.
The resolved signals will include a large number of bright galactic binaries, along with multiple supermassive black holes and EMRIs. 

The parameter estimation for the resolved systems will be impacted by the unresolved signals, which add to the effective noise level, and by the other resolved signals due to signal overlap. To simplify the discussion imagine that the resolved signals consist of bright galactic binaries $\textbf{H}_G$ and a single massive black hole binary $\bar{\textbf{h}}$. The parameter vector $\vec{\lambda}$ runs over the full set of galactic binary parameters (denoted by indices early in the alphabet, $\lambda^a,\lambda^b\dots$ {\it etc}) and the black hole parameters (denoted by indices later in the alphabet $\lambda^i,\lambda^j\dots$ {\it etc}). The full set of signal parameters for the galactic sources and the black hole are indicated by Greek indices. The Fisher information matrix for the combined solution,
$\Gamma_{\alpha\beta}$ can be broken into a block diagonal part $B_{\alpha\beta}$ formed from a galactic-binary block, $G_{a b}=(\partial_{a}\textbf{H}_G | \partial_{b}\textbf{H}_G)$, and a black hole
block $\bar{B}_{ij}=(\partial_{i} \bar{\textbf{h}} | \partial_{j} \bar{\textbf{h}})$, and a mixed block $M_{aj}= (\partial_{a}\textbf{H}_G | \partial_{j} \bar{\textbf{h}})$. The waveform error for the black hole signal is then
\begin{equation}
\Delta \bar{\textbf{h}} = -\partial_{i} \bar{\textbf{h}} \left(\textbf{n}|\partial_{\alpha}\textbf{H}_R\right)\left(\Gamma^{-1}\right)^{i \alpha} + \ldots \,\, .
\end{equation}
The expectation value for the squared SNR of the black hole waveform residual is then
\begin{equation}
\mathbbm{E}\left[ \left(\Delta \bar{\textbf{h}}|\Delta \bar{\textbf{h}}\right) \right] = \bar{B}_{ij} \left(\Gamma^{-1}\right)^{ij} + \ldots \,\, ,
\end{equation}
Assuming the mixture terms $M_{aj}$ are small compared to the terms on the diagonal, the inverse of the full Fisher matrix can be expanded as
\begin{eqnarray}
\left(\Gamma^{-1}\right)^{\alpha \beta} = &\left(B^{-1}\right)^{\alpha \beta}  - \left(B^{-1}\right)^{\alpha \mu} M_{\mu \nu}\left(B^{-1}\right)^{\nu \beta} \nonumber\\
&+  \left(B^{-1}\right)^{\alpha \mu}M_{\mu \nu}\left(B^{-1}\right)^{\nu \gamma}M_{\gamma \eta}\left(B^{-1}\right)^{\eta \beta} + \ldots \,\, .
\label{eq:mixed_inv_fisher}
\end{eqnarray}
The black-hole block of the inverse, $\left(\Gamma^{-1}\right)^{ij}$, lacks the second term since $\left(B^{-1}\right)^{i \mu} M_{\mu \nu}\left(B^{-1}\right)^{\nu j} = \left(\bar{B}^{-1}\right)^{i k} M_{k l}\left(\bar{B}^{-1}\right)^{l j} = 0$.
Therefore we have
\begin{eqnarray}
\mathbbm{E}\left[ \left(\Delta \bar{\textbf{h}}|\Delta \bar{\textbf{h}}\right) \right] &=  \bar{B}_{ij} \left(  \left(\bar{B}^{-1}\right)^{ij} + \left(\bar{B}^{-1}\right)^{ik}M_{k a} \left(G^{-1}\right)^{ab} M_{a l} \left( \bar{B}^{-1}\right)^{lj} \right) + \dots \nonumber \\
& =  \bar{D} + \left(\bar{B}^{-1}\right)^{ij}\left(G^{-1}\right)^{ab}M_{j a}M_{b i} + \ldots \,\, .
\end{eqnarray}
where $\bar{D}$ is the dimension of the black hole model. We see that the black hole waveform residuals are inflated from the isolated source result by the mixture term
\begin{eqnarray}
\left(\bar{B}^{-1}\right)^{ij}\left(G^{-1}\right)^{ab}M_{j a}M_{b i} &= \mathbbm{E}\left[ \Delta \lambda^{i}\Delta \lambda^{j} \right]\mathbbm{E}\left[ \Delta \lambda^{a}\Delta \lambda^{b} \right] \left(\partial_{j}\bar{\textbf{h}}|\partial_{a}\textbf{H}_{G}\right)\left(\partial_{b}\textbf{H}_{G}|\partial_{i}\bar{\textbf{h}}\right) \nonumber \\ 
&\approx \mathbbm{E} \left[ \Delta \lambda^{i}\Delta \lambda^{j} \Delta \lambda^{a} \Delta \lambda^{b}\right]\left(\partial_{j}\bar{\textbf{h}}|\partial_{a}\textbf{H}_{G}\right)\left(\partial_{b}\textbf{H}_{G}|\partial_{i}\bar{\textbf{h}}\right) \nonumber \\
&=  \mathbbm{E}\left [\left(\Delta \bar{\textbf{h}}|\Delta \textbf{H}_{G}\right)^2\right] \,\, .
\end{eqnarray}
The second line is obtained using Isserilis's theorem and dropping the cross terms $\mathbbm{E} [ \Delta \lambda^{i} \Delta \lambda^{a} ]$ which would produce higher order corrections. Using
$ \mathbbm{E}[ \Delta \tilde{H}_G(f)  \Delta \tilde{H}^*_G(f')]  = \frac{1}{2} S_{\Delta H_{G}}(f) \delta(f-f')$ we have
\begin{equation}
 \mathbbm{E}\left [\left(\Delta \bar{\textbf{h}}|\Delta \textbf{H}_{G}\right)^2\right] = 4 \int^{\infty}_{0} \frac{ |\Delta \bar{h}(f)|^2  S_{\Delta H_{G}}(f) }{S^2_{n,0}(f)} df \, .
\end{equation}
If we were to switch the roles of the black hole and the resolved galactic binaries we would find the the squared SNR of the galactic residuals were inflated by exactly the same amount:
\begin{equation}
\mathbbm{E}\left[ \left(\Delta {\textbf{H}_G}|\Delta {\textbf{H}_G}\right) \right] = D_G + \mathbbm{E}\left [\left(\Delta \bar{\textbf{h}}|\Delta \textbf{H}_{G}\right)^2\right] \, .
\end{equation}
On the other hand, the squared SNR of the full residual is equal to the total parameter dimension:
\begin{equation}
\mathbbm{E}\left[ \left(\Delta {\textbf{H}_R}|\Delta {\textbf{H}_R}\right) \right] = \Gamma_{\alpha \beta} \left(\Gamma^{-1}\right)^{\alpha \beta} = \bar{D} + D_G\, .
\end{equation}
These results can be reconciled by noting that
\begin{equation}
\fl \mathbbm{E}\left[ \left(\Delta {\textbf{H}_R}|\Delta {\textbf{H}_R}\right) \right] = \mathbbm{E}\left[ \left(\Delta {\textbf{H}_G}|\Delta {\textbf{H}_G}\right) \right] + \mathbbm{E}\left[ \left(\Delta \bar{\textbf{h}}|\Delta \bar{\textbf{h}}\right) \right] + 2 \mathbbm{E}\left[ \left( \Delta \bar{\textbf{h}} | \Delta \textbf{H}_G\right) \right]  \, ,
\end{equation}
and using
\begin{eqnarray}
\fl  \mathbbm{E}\left[ \left( \Delta \bar{\textbf{h}} | \Delta \textbf{H}_G\right) \right]  &= M_{ia} \Gamma_{\alpha \beta} \left(\Gamma^{-1}\right)^{i\alpha} \left(\Gamma^{-1}\right)^{a\beta} \nonumber \\
&=-M_{ia} M_{jb}  \left(\bar{B}^{-1}\right)^{ij} \left(G^{-1}\right)^{ab} \approx - \mathbbm{E}\left [\left(\Delta \bar{\textbf{h}}|\Delta \textbf{H}_{G}\right)^2\right] \, .
\end{eqnarray}
Thus we see that the extra residual for each source class is canceled by the cross-correlation of the residuals between the source classes. Note that the results for the SNR of the signal residuals are unchanged to the order we are considering when using the full noise model or the unperturbed noise model.

Next we consider the impact on the black hole parameter estimation errors caused by fitting the bright galactic binaries. The variance in the parameter estimation errors can be estimated from the diagonal entries of inverse of the full Fisher information matrix
\begin{equation}
\left(\Gamma^{-1}\right)^{ii} = \left(\bar{B}^{-1}\right)^{i i}   +  \left(\bar{B}^{-1}\right)^{i k}M_{k a}\left(G^{-1}\right)^{a b}M_{b n}\left(\bar{B}^{-1}\right)^{n i} + \ldots \,\, .
\label{eq:bhfish}
\end{equation}
The second term in the expansion comes from correlations between the black hole signal and the resolved galactic binaries as is positive definite since $\textbf{x}^{T} \mathbbm{G}\textbf{x}\geq 0$ for a positive-definite matrix. Thus, the simultaneous fitting of the galactic binary signals and the black hole signal tends to inflate the parameter estimation errors. Expanding to leading order the second term is given by
\begin{eqnarray}
\fl \bar{B}^{i k}M_{k a}G^{a b}M_{b n}\bar{B}^{n i} &= \mathbbm{E}\left[\Delta \lambda^{a} \Delta \lambda^{b} \left(\partial_{k}\bar{\textbf{h}}|\partial_{a}\textbf{H}_{R}\right)_0 \left(\partial_{b}\textbf{H}_{R}|\partial_{n}\bar{\textbf{h}}\right)_0\right]\left(\bar{B}_0^{-1}\right)^{i k}\left(\bar{B}_0^{-1}\right)^{n i}, \nonumber \\
&= \mathbbm{E}\left[ (\textbf{h}_{,k}|\Delta \textbf{H}_{G} )_0 (\Delta \textbf{H}_{G}|\textbf{h}_{,n})_0        \right] \left(\bar{B}_0^{-1}\right)^{i k}\left(\bar{B}_0^{-1}\right)^{n i} \nonumber \\
& =  4 \left(\bar{B}_0^{-1}\right)^{i k}\left(\bar{B}_0^{-1}\right)^{n i} \left(\mathbbm{R}   \int_0^\infty   \frac{ \partial_{k} \bar{\textbf{h}}^{*} \partial_{n} \bar{\textbf{h}} \, S_{\Delta H_{G}}}{S_{n,0}^2(f) } df\right) \, ,
\end{eqnarray}
where $\partial_{a}H_{R} = \partial_{a}H_{G}$ as the derivatives are with respect to GB parameters. Using an estimate to the waveform errors of the resolved sources we can express this result in a more useful form:

\begin{eqnarray}
4 \left(\mathbbm{R}   \int_0^\infty   \frac{ \partial_{k} \bar{\textbf{h}}^{*} \partial_{n} \bar{\textbf{h}} \, S_{\Delta H_{G}}}{S_{n,0}^2(f) } df\right) &\approx  2 D_{\scriptsize \mbox{eff}} \left(\mathbbm{R}   \int_0^\infty   \frac{ \partial_{k} \bar{\textbf{h}}^{*} \partial_{n} \bar{\textbf{h}} }{S_{n,0}(f)}\frac{dN}{db} df\right) \\
&\leq 2 D_{\scriptsize \mbox{eff}} \left(\mathbbm{R}   \int_0^\infty   \frac{ \partial_{k} \bar{\textbf{h}}^{*} \partial_{n} \bar{\textbf{h}} }{S_{n,0}(f)} df\right)\left(\frac{dN}{db}\right)_{\scriptsize\mbox{max}} \, .
\end{eqnarray}
This implies that the covariance matrix inflates with the following upper bound:

\begin{equation}
\left(\Gamma^{-1}\right)^{ii} \leq \bar{B}_{0}^{ii}\left( 1+2 D_{\scriptsize \mbox{eff}} \left(\frac{dN}{db}\right)_{\scriptsize\mbox{max}}\right) \, ,
\label{eq:cov_overlap_env}
\end{equation}
where $dN/db$ of sources resolved per frequency bin. In the next section we will obtain an estimate for the noise due to GB waveform errors which will allow us to obtain a more useful expression for this overlap term. A similar inflation of GB parameter variances results from the overlap with the BH signal.

\section{Galactic Confusion Noise}\label{sec:confusion}

\begin{figure}[htp]
	\begin{centering}
		\includegraphics[clip=true,angle=0,width=0.8\textwidth]{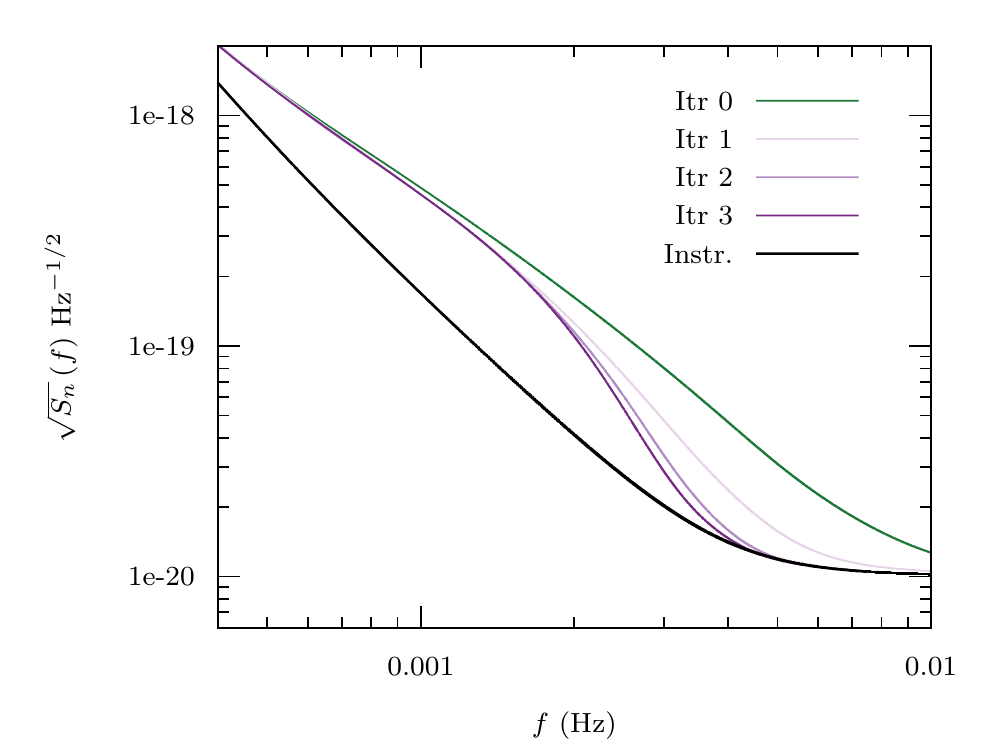} 
		\caption{\label{fig:iterate}  Estimates of the combined power spectral density of the combined instrument noise and galactic signals for the first 3 iterations on the removal process. These simulations are for a 4-year mission lifetime. For reference, the instrument noise contribution is shown as a heavy black line.}
	\end{centering}
\end{figure}

To obtain an estimate of the Galactic confusion foreground we employ an iterative subtraction scheme. Previously, this scheme was performed with perfect removal of source waveform~\cite{PhysRevD.73.122001, Cornish:2017vip}, but clearly noise will lead to errors in the parameters estimation and signal subtraction, and a reduction of the confusion noise estimate.

The revised procedure is as follows: A simulated data set is produced that includes a realization of the instrument noise and the sum of the strains due to the galactic binaries $H$ from the galactic population model. A smooth fit to the power spectral density of the instrument noise and the signals is used as an initial estimate for the noise in each data channel. Next we identify sources which are loud (SNR $> 7$) relative to this noise estimate and subtract the best-fit waveform $\textbf{h}(\vec{\lambda})_{{\rm best fit}} = \textbf{h}_T-\partial_i\textbf{h}_T\Delta \lambda^i $ from the data. A smooth fit to the power spectral density of the remaining signals and noise is computed, and signals above the SNR threshold for the updated noise estimate are identified and subtracted. As can be seen in Figure~\ref{fig:iterate}, the subtraction procedure quickly converges.

\begin{figure}[htp]
	\begin{centering}
		\includegraphics[clip=true,angle=0,width=0.8\textwidth]{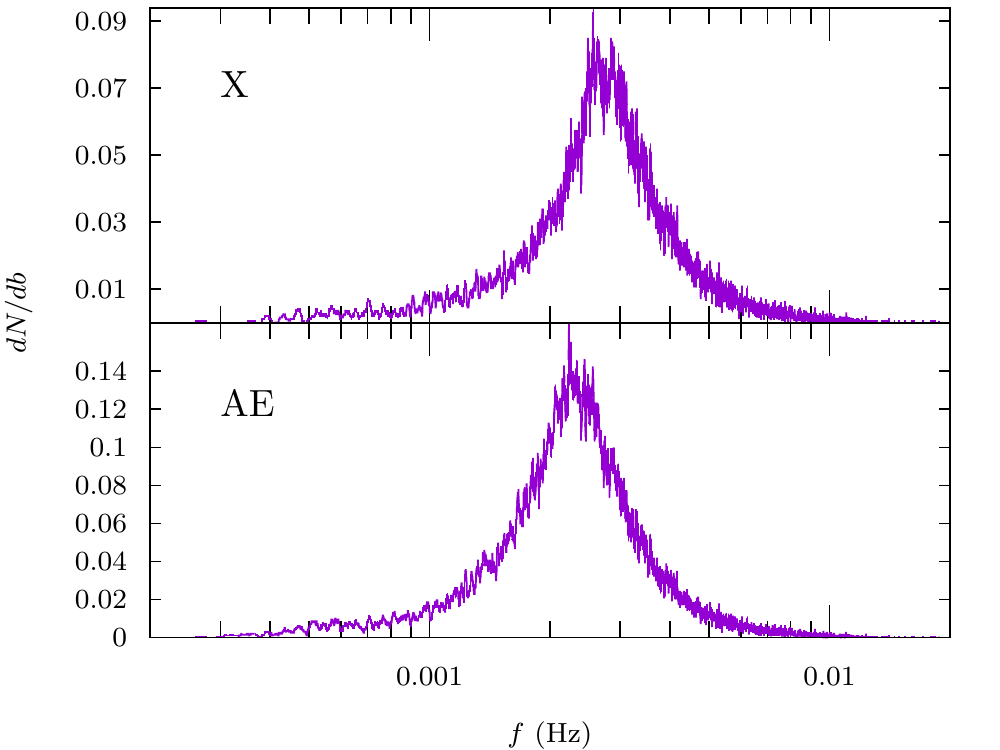} 
		\caption{\label{fig:dndb}  The number of sources resolved per frequency bin, $dN/db$, for a single channel analysis $X$, and a dual channel analysis $AE$, for a 4-year mission duration. Note the difference in the vertical scales; the combined information in the $AE$ data stream allows for more sources to be resolved. The density of resolvable systems peaks at around $2-3$ mHz.}
	\end{centering}
\end{figure}

The number of sources which can be resolved converges after just 5 or 6 iterations. The number density of sources, measured in terms of the number per a frequency bin, $dN/db$, is shown in Figure~\ref{fig:dndb} for a single channel and dual channel analysis. We see that more sources can be resolved when multiple data channels are used in the analysis. Attempts have been made to deal with the identification and subtraction of signals which overlap \cite{PhysRevD.67.103001,0264-9381-22-18-S04} and of how many sources per frequency band \cite{2003gr.qc.....4020C} can be resolved. Here we see that with two data channels and a 4-year mission the peak density is roughly one source resolved per ten frequency bins. In our simulations we made the simplifying assumption that the augment Fisher matrix for the galactic population is block diagonal. That is, we ignored correlations between galactic signals. This approximation is reasonable when $dN/db$ is small, but may be questionable in the highest density regions and for the occasional systems that happen to have high overlap. We will re-visit this complication in a future study, as the parameter estimation errors grow significantly for highly overlapping systems~\cite{Crowder:2004ca}.

\begin{figure}[htp]
	\begin{centering}
		\includegraphics[clip=true,angle=0,width=0.8\textwidth]{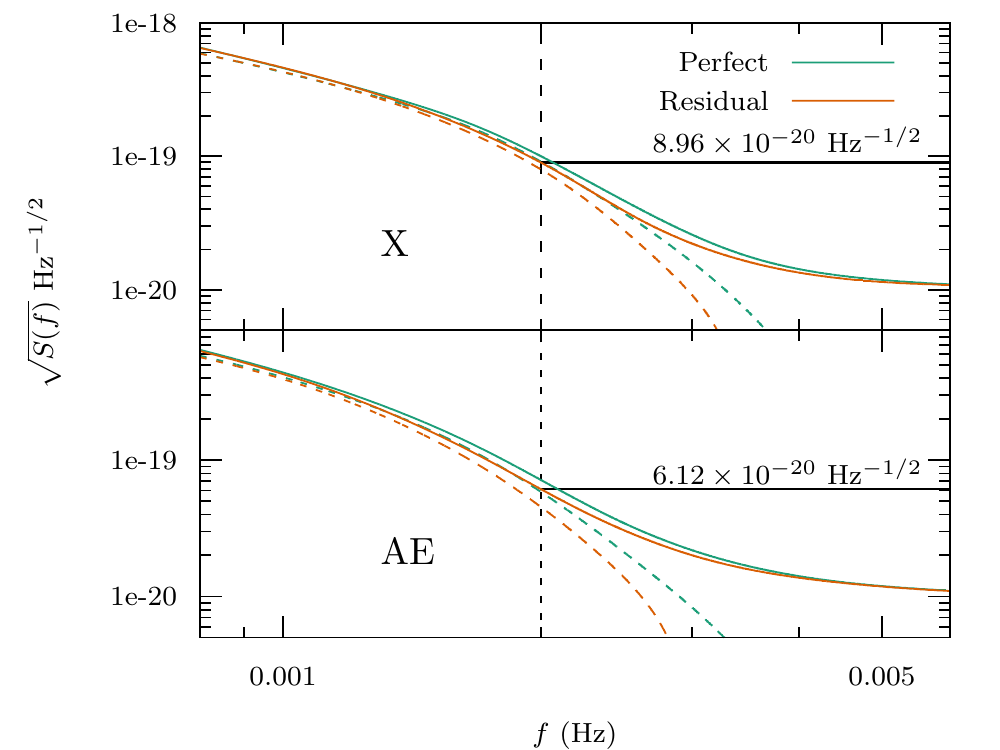} 
		\caption{\label{fig:perf_resid_comp} The Michelson-equivalent stain PSD in both the $X$ and combined $A, E$ data channels. The dashed lines represent the the GB confusion noise. A reference frequency of $2$ mHz has been denoted by a vertical dashed line for noise level comparison on the corresponding horizontal solid black lines. }
	\end{centering}
\end{figure}

Figure \ref{fig:perf_resid_comp} compares the Michelson-equivalent strain power spectral densities for the imperfect and perfect subtraction scheme. The dashed lines show galactic confusion  noise and the solid lines show the combined instrument and confusion noise.
Note that the differences between the PSD arise where the most sources are resolved as one would expect (see Figure \ref{fig:dndb}). In the dual $A, E$ channel the PSD is lower as indicated by the noise levels specified by the reference frequency $2$ mHz.

We can estimate the power spectral density of the combined waveform residual,  $S_{\Delta H_{R}}(f) = \frac{T}{2}\mathbbm{E}\left[ |\Delta \tilde{H}_{R}(f)|^{2}\right]$ by applying
(\ref{eq:dhsq}) to the full compliment of $N$ resolved binaries:
\begin{equation}
\mathbbm{E}\left[\rho_{\Delta H_{R}}^{2}\right] = N D_{\rm eff} = 4 \int_{0}^{\infty}\frac{\mathbbm{E}\left[|\Delta \tilde{H}_{R}(f)|^{2}\right]}{S_{n,0}(f)}df \nonumber  \,\, ,
\end{equation}
Considering the contribution in a small frequency range $\Delta f$ centered at $f$ we find
\begin{equation}
S_{\Delta H_{R}}(f) = \frac{D_{\rm eff}}{2} \frac{dN}{db} S_{n,0}(f) \,\, .
\label{eq:analytic_psd}
\end{equation}

\begin{figure}[htp]
	\begin{centering}
		\includegraphics[clip=true,angle=0,width=0.8\textwidth]{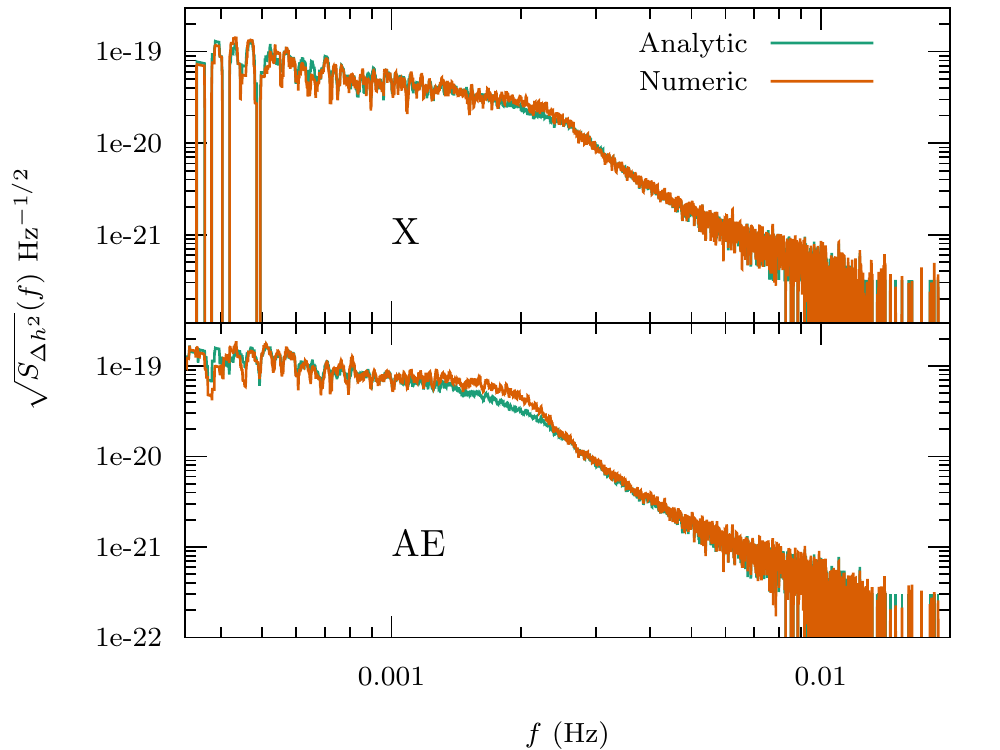} 
		\caption{\label{fig:analytic_numeric}  Comparison between the predicted (teal) strain power spectral density and that obtained through the imperfect subtraction scheme (orange) of the GB confusion strictly plotted as their Michelson-equivalent strain.}
	\end{centering}
\end{figure}
Figure \ref{fig:analytic_numeric} compares this analytic estimate to the numerical value found from the iterative subtraction scheme. The prediction lines up quite well with a small deviation at the frequencies where the number of resolved sources per bin peaks.

\section{Discussion}

We have used the maximum likelihood approximation to derive a number of analytic results pertaining to the LISA global analysis problem. A simple toy model was used to demonstrate the relevance of these estimates to a full Bayesian analysis, though we cannot guarantee that the approximations will be as reliable when applied to LISA data analysis. We extended the standard maximum likelihood analysis to include noise modeling, and found that the estimated noise level is lowered whenever signals are subtracted from the data. We applied our general results to the simultaneous fitting of a black hole binary and a collection of galactic binaries and found that the errors in the black hole waveform recovery are increased, as wells as the variances of the source parameters. It is important to note that it is not the overlap of the signals which cause the inflation of variances, but rather, it is due to the waveform subtraction errors. We concluded by incorporating parameter estimation errors in the estimation of the galactic confusion noise, and derived a useful expression that can be used to predict the reduction in the confusion noise in terms of the number density of resolved signals. Equations (\ref{eq:cov_envo}) and (\ref{eq:cov_overlap_env}) provide quick estimates for how parameter estimation errors are inflated by noise fitting and source confusion.

\subsection*{Acknowledgments}
We are grateful for the support provided by NASA grant NNX16AB98G.

\section*{Appendix A}
Here we derive the more general signal plus noise model ML analysis results presented in section 4. Beginning with the log-likelihood obtained from equation (\ref{eq:likelihood})

\begin{eqnarray} 
\log p\left(\textbf{s}\Big|\textbf{h}(\vec{\lambda}),S_{n}(f;\vec{\theta})\right) = &-\frac{1}{2}\int_{0}^{\infty} T \log\left[\pi T S_{n}(f;\vec{\theta})\right]df \nonumber \\
&- \frac{1}{2}\left(\textbf{s}-\textbf{h}(\vec{\lambda})\Big|\textbf{s}-\textbf{h}(\vec{\lambda})\right)_{\vec{\theta}}  \,\, ,
\end{eqnarray}
where the noise-weighted inner product is now parameterized by $\vec{\theta}$ as denoted by $( \cdot | \cdot )_{\vec{\theta}}$. We may expand about the true noise model $S_{n,0}(f)$ and the true signal model $\textbf{h}_{T}$ and maximized to obtain estimates of  $\Delta \vec{\theta}$ and $\Delta \vec{\lambda}$. Expanding out the normalization constant and dropping terms which are constant with respect the maximization gives

\begin{eqnarray}
-\frac{1}{2}\int_{0}^{\infty} \log\left[2\pi T S_{n}(f;\vec{\theta})\right]df \approx& -\frac{1}{2}\Delta\theta^{i}\int_{0}^{\infty}\frac{T S_{n,i}}{S_{n,0}}df-\frac{1}{4}\Delta\theta^{i}\Delta\theta^{j}\int_{0}^{\infty}\frac{T S_{n,ij}}{S_{n,0}}df  \nonumber \\ &+\frac{1}{4}\Delta\theta^{i}\Delta\theta^{j}\int_{0}^{\infty}\frac{T S_{n,i}S_{n,j}}{S_{n,0}^{2}}df  \,\, ,
\end{eqnarray}
where $T$ is the observation period. An arbitrary noise-weighted inner product expanded out takes the form

\begin{eqnarray}
(\textbf{a}|\textbf{b})_{\vec{\theta}} \approx (\textbf{a}|\textbf{b})_{0} -\Delta\theta^{i}(\textbf{a}|\textbf{b})_{i} - \frac{1}{2}\Delta\theta^{i}\Delta\theta^{j}(\textbf{a}|\textbf{b})_{ij} +\Delta\theta^{i}\Delta\theta^{j}(\textbf{a}|\textbf{b})_{i;j} \,\, ,
\end{eqnarray}
where 
\begin{eqnarray}
(\textbf{a}|\textbf{b})_{i} &\defeq 4 \mathbbm{R}\int_{0}^{\infty} \frac{\tilde{a}^{*}\tilde{b}}{S_{n,0}}\frac{S_{n,i}}{S_{n,0}}df \nonumber \,\, , \\
(\textbf{a}|\textbf{b})_{ij} &\defeq 4 \mathbbm{R}\int_{0}^{\infty} \frac{\tilde{a}^{*}\tilde{b}}{S_{n,0}}\frac{S_{n,ij}}{S_{n,0}}df \nonumber \,\, , \\
(\textbf{a}|\textbf{b})_{i;j} &\defeq 4 \mathbbm{R}\int_{0}^{\infty} \frac{\tilde{a}^{*}\tilde{b}}{S_{n,0}}\frac{S_{n,i}S_{n,j}}{S_{n,0}^{2}}df \nonumber  \,\, .
\end{eqnarray}
With these pieces in hand the chi-squared piece of the log-likelihood, dropping constants with respect to the maximization, is
\begin{eqnarray}
- \frac{1}{2}\left(\textbf{s}-\textbf{h}(\vec{\lambda})\Big|\textbf{s}-\textbf{h}(\vec{\lambda})\right)_{\vec{\theta}} \approx& \frac{1}{2}\Delta \theta^{i}(\textbf{n}|\textbf{n})_{i}+\frac{1}{4}\Delta \theta^{i}\Delta \theta^{j}(\textbf{n}|\textbf{n})_{ij} \nonumber \\
&-\frac{1}{2}\Delta\theta^{i}\Delta\theta^{j}(\textbf{n}|\textbf{n})_{i;j}+\Delta \theta^{i}(\textbf{n}|\Delta \textbf{h})_{i} \nonumber \\
&+\frac{1}{2}(\textbf{n}|\Delta \textbf{h})_{0}+\frac{1}{2}\Delta \theta^{i} (\Delta \textbf{h}|\Delta\textbf{h})_{i} \nonumber \\
&-\frac{1}{2}(\Delta \textbf{h}|\Delta\textbf{h})_{0}
\end{eqnarray}
Collecting terms, and maximizing results in the solution
\begin{eqnarray}
\Delta \lambda^{j} & = (\textbf{n}|\partial_{j}\textbf{h}_{T})_{\vec{\theta}}\left(\Gamma_{\vec{\theta}}^{-1}\right)^{ij}\\
\Delta \theta^{j} \approx &\left[ \int \frac{S_{n,i}S_{n,j}}{S_{n,0}^{2}}\left(\frac{T S_{n,0}- 4 \tilde{n}^{*}\tilde{n}}{S_{n,0}}\right)df - \int \frac{S_{n,ij}}{S_{n,0}}\left(\frac{T S_{n,0}- 2 \tilde{n}^{*}\tilde{n}}{S_{n,0}}\right)df \right]^{-1} \nonumber \\
& \times \left[\int \frac{S_{n,i}}{S_{n,0}}\left(\frac{T S_{n,0}- 2 \tilde{n}^{*}\tilde{n}}{S_{n,0}}\right)df - 2(\textbf{n}|\Delta\textbf{h})_{i} - (\Delta\textbf{h}|\Delta\textbf{h})_{i} \right] \,\, ,
\end{eqnarray}
where $\Delta \textbf{h}$ need only be kept to leading order. The Fisher matrix can be obtained similarly to the toy white noise problem presented in section \ref{sec:MLnoise}

\begin{equation}
\boldsymbol{\Gamma} = \left(
\begin{array}{cc}
	(\partial_{i}\textbf{h}_{T}|\partial_{j}\textbf{h}_{T})_{0} + \frac{1}{2}(\textbf{n}|\partial_{ij}\textbf{h}_{T})_{0} & (\textbf{n}|\partial_{j}\textbf{h}_{T})_{i} \\
	&\\
	(\textbf{n}|\partial_{i}\textbf{h}_{T})_{j} & \frac{1}{2} \int \frac{S_{n,ij}}{S_{n,0}}\left(\frac{ 2 \tilde{n}^{*}\tilde{n}-T S_{n,0}}{S_{n,0}}\right)df \\
	&+\frac{1}{2}\int \frac{S_{n,i}S_{n,j}}{S_{n,0}^{2}}\left( \frac{4 \tilde{n}^{*}\tilde{n} - TS_{n,0}}{S_{n,0}} \right) df
\end{array} \right) \, .
\end{equation}
Inverting the Fisher matrix and considering the signal model variances

\begin{eqnarray}
\left(\Gamma^{-1}\right)^{ii} \approx& \left(\Gamma^{-1}_{0}\right)^{ii} -\frac{1}{2}\left(\Gamma^{-1}_{0}\right)^{im} (\textbf{n}|\partial_{mn}\textbf{h}_{T})_{0}\left(\Gamma^{-1}_{0}\right)^{ni} \nonumber \\
&+ 2 \left(\Gamma^{-1}_{0}\right)^{im} (\textbf{n}|\partial_{m}\textbf{h}_{T})_{a} \left[\int_{0}^{\infty}\frac{S_{n,ab}}{S_{n,0}}\left(\frac{ 2 \tilde{n}^{*}\tilde{n}-T S_{n,0}}{S_{n,0}}\right)df \right. \nonumber \\
&\left. + \int_{0}^{\infty} \frac{S_{n,a}S_{n,b}}{S_{n,0}^{2}}\left( \frac{4 \tilde{n}^{*}\tilde{n} - TS_{n,0}}{S_{n,0}} \right) df\right]^{-1}(\textbf{n}|\partial_{n}\textbf{h}_{T})_{b}\left(\Gamma^{-1}_{0}\right)^{ni} \, .
\end{eqnarray}

\section{References}

\bibliography{bibliography}

\providecommand{\newblock}{}
\begin{thebibliography}{10}
\expandafter\ifx\csname url\endcsname\relax
  \def\url#1{{\tt #1}}\fi
\expandafter\ifx\csname urlprefix\endcsname\relax\def\urlprefix{URL }\fi
\providecommand{\eprint}[2][]{\url{#2}}

\bibitem{PhysRevLett.116.061102}
Abbott B~P {\em et~al.\/} (LIGO Scientific Collaboration and Virgo
  Collaboration) 2016 {\em Phys. Rev. Lett.\/} {\bf 116}(6) 061102
  \urlprefix\url{https://link.aps.org/doi/10.1103/PhysRevLett.116.061102}

\bibitem{PhysRevLett.116.231101}
Armano M {\em et~al.\/} 2016 {\em Phys. Rev. Lett.\/} {\bf 116}(23) 231101
  \urlprefix\url{https://link.aps.org/doi/10.1103/PhysRevLett.116.231101}

\bibitem{LISA16}
Danzmann K~{\it et al} 2016 Laser interferometer space antenna Tech. rep. LISA
  Consortium
  \urlprefix\url{https://www.elisascience.org/files/publications/LISA\_L3\_201%
70120.pdf}

\bibitem{Arnaud:2006gm}
Arnaud K~A {\em et~al.\/} 2006 {\em AIP Conf. Proc.\/} {\bf 873} 619--624
  [,619(2006)] (\textit{Preprint} \eprint{gr-qc/0609105})

\bibitem{Arnaud:2007vr}
Arnaud K~A {\em et~al.\/} 2007 {\em Class. Quant. Grav.\/} {\bf 24} S529--S540
  (\textit{Preprint} \eprint{gr-qc/0701139})

\bibitem{Arnaud:2007jy}
Arnaud K~A {\em et~al.\/} 2007 {\em Class. Quant. Grav.\/} {\bf 24} S551--S564
  (\textit{Preprint} \eprint{gr-qc/0701170})

\bibitem{Babak:2007zd}
Babak S {\em et~al.\/} (Mock LISA Data Challenge Task Force) 2008 {\em Class.
  Quant. Grav.\/} {\bf 25} 114037 (\textit{Preprint} \eprint{0711.2667})

\bibitem{Babak:2008aa}
Babak S {\em et~al.\/} 2008 {\em Class. Quant. Grav.\/} {\bf 25} 184026
  (\textit{Preprint} \eprint{0806.2110})

\bibitem{Babak:2009cj}
Babak S {\em et~al.\/} (Mock LISA Data Challenge Task Force) 2010 {\em Class.
  Quant. Grav.\/} {\bf 27} 084009 (\textit{Preprint} \eprint{0912.0548})

\bibitem{PhysRevD.73.122001}
Timpano S~E, Rubbo L~J and Cornish N~J 2006 {\em Phys. Rev. D\/} {\bf 73}(12)
  122001 \urlprefix\url{http://link.aps.org/doi/10.1103/PhysRevD.73.122001}

\bibitem{Cornish:2017vip}
Cornish N and Robson T 2017 {Galactic binary science with the new LISA design}
  {\em {11th International LISA Symposium Zurich, Switzerland, September 5-9,
  2016}\/} (\textit{Preprint} \eprint{1703.09858})
  \urlprefix\url{http://inspirehep.net/record/1519956/files/arXiv:1703.09858.p%
df}

\bibitem{PhysRevD.73.042001}
Cutler C and Harms J 2006 {\em Phys. Rev. D\/} {\bf 73}(4) 042001
  \urlprefix\url{https://link.aps.org/doi/10.1103/PhysRevD.73.042001}

\bibitem{Crowder:2004ca}
Crowder J and Cornish N~J 2004 {\em Phys. Rev.\/} {\bf D70} 082004
  (\textit{Preprint} \eprint{gr-qc/0404129})

\bibitem{0004-637X-758-2-131}
Nissanke S, Vallisneri M, Nelemans G and Prince T~A 2012 {\em The Astrophysical
  Journal\/} {\bf 758} 131
  \urlprefix\url{http://stacks.iop.org/0004-637X/758/i=2/a=131}

\bibitem{2007MNRAS.382..685R}
{Roelofs} G~H~A, {Nelemans} G and {Groot} P~J 2007 {\em \mnras\/} {\bf 382}
  685--692 (\textit{Preprint} \eprint{0709.2951})

\bibitem{2013MNRAS.429.2143C}
{Carter} P~J, {Marsh} T~R, {Steeghs} D, {Groot} P~J, {Nelemans} G, {Levitan} D,
  {Rau} A, {Copperwheat} C~M, {Kupfer} T and {Roelofs} G~H~A 2013 {\em
  \mnras\/} {\bf 429} 2143--2160 (\textit{Preprint} \eprint{1211.6439})

\bibitem{1990ApJ...360...75H}
{Hils} D, {Bender} P~L and {Webbink} R~F 1990 {\em \apj\/} {\bf 360} 75--94

\bibitem{0004-637X-537-1-334}
Hils D and Bender P~L 2000 {\em The Astrophysical Journal\/} {\bf 537} 334
  \urlprefix\url{http://stacks.iop.org/0004-637X/537/i=1/a=334}

\bibitem{PhysRevD.84.063009}
Littenberg T~B 2011 {\em Phys. Rev. D\/} {\bf 84}(6) 063009
  \urlprefix\url{https://link.aps.org/doi/10.1103/PhysRevD.84.063009}

\bibitem{2012A&A...546A..70T}
{Toonen} S, {Nelemans} G and {Portegies Zwart} S 2012 {\em \aap\/} {\bf 546}
  A70 (\textit{Preprint} \eprint{1208.6446})

\bibitem{PhysRevD.76.083006}
Cornish N~J and Littenberg T~B 2007 {\em Phys. Rev. D\/} {\bf 76}(8) 083006
  \urlprefix\url{https://link.aps.org/doi/10.1103/PhysRevD.76.083006}

\bibitem{0004-637X-527-2-814}
Armstrong J~W, Estabrook F~B and Tinto M 1999 {\em The Astrophysical Journal\/}
  {\bf 527} 814 \urlprefix\url{http://stacks.iop.org/0004-637X/527/i=2/a=814}

\bibitem{Prince:2002hp}
Prince T~A, Tinto M, Larson S~L and Armstrong J~W 2002 {\em Phys. Rev.\/} {\bf
  D66} 122002 (\textit{Preprint} \eprint{gr-qc/0209039})

\bibitem{PhysRevD.77.042001}
Vallisneri M 2008 {\em Phys. Rev. D\/} {\bf 77}(4) 042001
  \urlprefix\url{https://link.aps.org/doi/10.1103/PhysRevD.77.042001}

\bibitem{PhysRevD.49.2658}
Cutler C and Flanagan {\'{E}}~E 1994 {\em Phys. Rev. D\/} {\bf 49} 2658--2697
  \urlprefix\url{https://link.aps.org/doi/10.1103/PhysRevD.49.2658}

\bibitem{0004-637X-575-2-1030}
Takahashi R and Seto N 2002 {\em The Astrophysical Journal\/} {\bf 575} 1030
  \urlprefix\url{http://stacks.iop.org/0004-637X/575/i=2/a=1030}

\bibitem{Adams:2012qw}
Adams M~R, Cornish N~J and Littenberg T~B 2012 {\em Phys. Rev.\/} {\bf D86}
  124032 (\textit{Preprint} \eprint{1209.6286})

\bibitem{Littenberg:2014oda}
Littenberg T~B and Cornish N~J 2015 {\em Phys. Rev.\/} {\bf D91} 084034
  (\textit{Preprint} \eprint{1410.3852})

\bibitem{PhysRevD.67.103001}
Cornish N~J and Larson S~L 2003 {\em Phys. Rev. D\/} {\bf 67}(10) 103001
  \urlprefix\url{https://link.aps.org/doi/10.1103/PhysRevD.67.103001}

\bibitem{0264-9381-22-18-S04}
Umstätter R, Christensen N, Hendry M, Meyer R, Simha V, Veitch J, Vigeland S
  and Woan G 2005 {\em Classical and Quantum Gravity\/} {\bf 22} S901
  \urlprefix\url{http://stacks.iop.org/0264-9381/22/i=18/a=S04}

\bibitem{2003gr.qc.....4020C}
{Cornish} N~J 2003 {\em ArXiv General Relativity and Quantum Cosmology
  e-prints\/} (\textit{Preprint} \eprint{gr-qc/0304020})

\end{thebibliography}

\end{document}